\documentclass[11pt]{article}

\usepackage{fullpage}

\usepackage{amsmath,amsfonts, amssymb,amsthm}              
\usepackage{dsfont}
\usepackage{graphics, epsfig}
\usepackage{amsfonts}
\usepackage{graphics,color}
\usepackage{algorithm, algorithmic}
\usepackage{appendix}
\usepackage{enumerate}
\usepackage{multirow}
\usepackage[table]{xcolor}
\usepackage{hyperref}

\usepackage{graphicx}
\usepackage{epsfig}

\usepackage[lofdepth,lotdepth]{subfig}
\usepackage{caption}

\usepackage{float}

\floatstyle{boxed}
\newfloat{algorithm}{thp}{lop}
\floatname{algorithm}{Algorithm}
\usepackage{authblk}
\providecommand{\keywords}[1]{\textbf{\textit{Keywords---}} #1}
\begin{document}
\graphicspath{{./NewFigs/}}

\title{In the mood: the dynamics of collective sentiments on Twitter}
\author[1,2]{Nathaniel Charlton\thanks{billiejoecharlton@gmail.com}}
\author[1,2]{Colin Singleton\thanks{c.singleton@reading.ac.uk}}
\author[2]{Danica Vukadinovi\'c Greetham \thanks{d.v.greetham@reading.ac.uk}}
\affil[1]{CountingLab Ltd, Reading, UK}
\affil[2]{Centre for the Mathematics of Human Behaviour\\
Department of Mathematics and Statistics\\
University of Reading, UK}

\date{March 7, 2016}
\maketitle              
\begin{abstract}
We study the relationship between the sentiment levels of Twitter users and the evolving network structure that the users created by @-mentioning each other.
We use a large dataset of tweets to which we apply three sentiment scoring algorithms, including the open source SentiStrength program.
Specifically we make three contributions.
Firstly we find that people who have potentially the largest communication reach (according to a dynamic centrality measure) use sentiment differently than the average user:
for example they use positive sentiment more often and negative sentiment less often.
Secondly we find that when we follow structurally stable Twitter communities over a period of months, their sentiment levels are also stable, and sudden changes in community sentiment from one day to the next can in most cases be traced to external events affecting the community.
Thirdly, based on our findings, we create and calibrate a simple agent-based model that is capable of reproducing measures of emotive response comparable to those obtained from our empirical dataset.
\end{abstract}

\maketitle
\section{Introduction}\label{intro}
It has been noticed long before the Internet that emotions appear to be contagious (see \cite{Doherty:97}). While different mechanisms were proposed to explain this phenomenon,
from complex cognitive  processes \cite{Aronfreed:70}, to automatic mimicry and
synchronisation of facial, vocal,
postural, and instrumental expressions with those around us \cite{Hatfield:92}, it is not yet clear how reverberating or inhibiting is online social media regarding contagion of emotions.
Agent-based modelling was used to model dynamics of sentiments in online forums\cite{Garcia:10, Tadic:13}
and to look at the recent rise of the 15M movement in Spain \cite{Moreno:15}. It has been shown in \cite{Greetham:11} that positive and negative affects \cite{Tellegen:1985} that are sometimes used to describe positive and negative mood are not complementary and follow different dynamics in a social network. Furthermore, it was conjectured in \cite{Greetham:15} that the people with the potentially largest reach to all the others  in a smaller social network over a week belong to the group with the smallest negative affect at the beginning of that period. In this work we investigate whether similar conclusions can be discovered for large online social networks, using automatic sentiment detection algorithms, and to what extent we can develop a good model of collective sentiments dynamics.
Our contributions are threefold:
\begin{itemize}
\item Firstly, we apply \emph{dynamic communicability}, a centrality measure for evolving networks, to a snowball-sampled Twitter network, allowing us to identify the ``top broadcasters''
i.e.\ those users with potentially the highest communication reach in the network. We find that people with the highest communicability broadcast indices show different patterns of sentiment use compared to ordinary users. For example, top broadcasters send positive sentiment messages more often, and negative sentiment messages less often. When they do use positive sentiment, it tends to be stronger.

\item Secondly, by using a number of community detection algorithms in combination, we were able to identify and monitor structurally stable (over a time-scale of months) ``communities'' or ``sub-networks'' of Twitter users. Users within these communities are well-connected and send messages to each other frequently compared with how frequently they send messages to users not in the community. We find that each such community has its own sentiment level, which is also relatively stable over time. We find that when the sentiment in a community temporarily shows a large deviation from its usual level, this can typically be traced to a significant identifiable event affecting the community, sometimes an external news event. Some of the communities we followed retained all their users over the period of monitoring, but the others lost a varying (but relatively small) proportion of their users. We find correlations between the loss of users and the conductance and initial sentiment of the communities.

\item Finally, an Agent-Based Model of online social networks is presented. The model consists of a population of simulated users, each with their own individual characteristics, such as their tendency to initiate new conversations, their tendency to reply when they have been sent a message, and their usual sentiment level. The model allows for sentiment contagion, where users' sentiment levels change in response to the sentiment of the messages they receive. We demonstrate that this model, when its parameters are fitted to data from a real Twitter community, accurately reproduces various aspects of that community.
\end{itemize}

Appendix~\ref{app:data_made_available} describes the data we can make available to support this article.

\section{Data}
\label{sec:data}

The data analysed in this work consists of posts (``tweets'') from Twitter. Twitter provides a platform for users to post short texts (up to 140 characters in length) for viewing by other users. Twitter users often direct or address their public tweets to other users by using mentions with the @ symbol. Suppose there are two users with usernames Alice and Bob. Alice might greet Bob by tweeting: ``@Bob Good morning, how are you today?''. Bob might reply with ``I am feeling splendid @Alice''. Note that although mentions are used to address other users in a tweet, the tweet itself is still public and the messages may be read and commented on by other users.

We comissioned a digital marketing agency to collect Twitter data for our experiments. This was done in two stages:

\begin{enumerate}

\item {\bf Snowball sampling of a large set of users.} We began with a single seed user. For the seed user, and each time we added a user to our sample, we retrieved that user's last 200 public tweets (or all their tweets if they had posted fewer than 200 since account creation), and identified other users they had mentioned.  These users were then added to the sample, and so on. In this manner 669,191 users were sampled and a total of 121,805,832 tweets collected. Limiting the history collected to the last 200 tweets enabled us to explore a larger subgraph of the Twitter network, and ensured that we would be able to find sufficiently many interesting communities for study. Informally speaking, our snowballed dataset was broad at the expense of being shallow.

\item {\bf Obtaining a detailed tweet history for selected interesting groups of users.} Once we had identified interesting communities of users for study (as we will describe in
Section~4.\ref{sub:community_detection}), containing altogether 10,000 distinct users,
we retreived a detailed tweet history for these users. We downloaded each user's pervious 3,200 tweets (a limit imposed by the Twitter API) obtaining altogether 22,469,713 tweets.
Note that the period covered by 3,200 tweets varies considerably depending on the tweeting frequency of the user: heavy users may post 3,200 tweets in just a few days, whereas for some light users 3,200 tweets extended all the way back to the year 2006. We also monitored the users ``live'' for a further period of 30 days, logging all their tweets posted during this time, yielding a further 3,216,136 tweets.
Informally speaking, this part of our dataset was deep (but at the expense of being narrower).

\end{enumerate}

Using sentiment analysis programs, we assigned three sentiment measures to each tweet, named and described as follows:

\begin{description}

\item{(MC)} This sentiment score was provided by the marketing company's highly tuned proprietary algorithm. The algorithm involves recognising words and phrases that typically indicate positive or negative sentiment,
but its exact details are not published, as it is commercial IP. The score for each tweet is an integer ranging from $-25$ (extremely negative) through 0 (neutral) up to +25 (extremely positive).

\item{(SS)} This sentiment score was produced by SentiStrength\footnote{http://sentistrength.wlv.ac.uk/} program \cite{Thelwall:2012}.
SentiStrength provides separate measures of the positive and negative sentiment of each tweet; we derive a single measure analogous to (MC) by subtracting the strength of the negative sentiment from the strength of the positive sentiment.
The (SS) measure ranges from $-4$ to +4.

\item{(L)} This sentiment score was produced by the LIWC2007\footnote{http://www.liwc.net/} program \cite{pennebaker:2007}. Like SentiStrength, LIWC produces separate measures {\it posemo} and {\it negemo} of positive and negative emotion; by subtracting {\it negemo} from {\it posemo}
we derived a single real-valued measure ranging from $-100$ to +100.

\end{description}

Although all three sentiment classifiers are the result of extensive development effort, none of them is perfect; this is to be expected given the subtlety of human language. Thus we think of the sentiment as a very ``noisy'' signal. The work described in this paper takes averages over large numbers of tweets and users, so our results do not depend on the exact score of particular individual tweets; we require only that on average the sentiment scores reflect the kind of sentiments expressed by users.

\section{Communicability and sentiment}
\label{sec:communicability_and_sent}

In this section we investigate how users with the highest potential communication reach tend to use sentiment in their messages.
We use \emph{dynamic communicability}, a centrality measure for evolving networks, to assign \emph{broadcast scores} to users; these scores are one method of quantifying communication reach that has been investigated in the literature.
Our investigation is motivated by the finding, in three small observed social network studies \cite{Greetham:15}, that the individuals with large broadcast scores in general had very low levels of negative affect  at the beginning of the studies.

\subsection{Broadcast scores}

In this subsection we briefly describe the measure we used to quantify potential communication reach. The measure, called \emph{dynamic communicability} \cite{Grindrod:2011}, is a centrality measure for evolving networks based on Katz centrality \cite{Katz:1953}. Katz centrality in static networks counts all possible paths from and to each vertex, penalising progressively longer paths. Let an evolving network be represented by a sequence of adjacency matrices $A_t$, where $t =1,\ldots,n$ is the time-step. Then dynamic communicability counts all the possible time-respecting paths over the evolving network: such a path can make for example one hop at time-step $t = 1$ and the next hop at time-step $t = 3$, but not vice versa. The formal definition we use for a dynamic communicability matrix is
\begin{displaymath}
Q=\prod_{t = 1}^n(I-\alpha A_t)^{-1},
\end{displaymath}
where $I$ is identity matrix, $\alpha < (\rho(A_{t}))^{-1}$ is a penalising factor  and $\rho (A_t)$ is the largest eigenvalue\footnote{Note that $\alpha < (\rho(A_{t}))^{-1}, \forall t = 1, \ldots, n$, for the inverse to exist. For computational reasons, $(I-\alpha A_t)^{-1} =\sum_{i=0}^\infty \alpha^iA_t^i$  is often approximated with a finite truncation of the $i$ first summands, which then also allows to relax the penalising constraint to $\alpha < 1$. } of $A_t$.
When $\alpha$ is small, short paths in the network are valued highly relative to long paths; when $\alpha$ is larger, long paths are given a relatively larger weight. Here we use
one ``snapshot'' $A_t$ for each day.

$Q$ is a square matrix, with rows and columns representing vertices or individuals in the network. The $k^{th}$ row and column sums each represent a measure of communicability for the vertex (user) $k$. The row sum represents the \emph{broadcast} index while the column sum measures the \emph{receive} index. As the respective names suggest, they measure how well the vertex $k$ is able to broadcast and receive messages over the network.

\subsection{Extracting a ``mentions'' network to analyse broadcast scores}
\label{sub:extracting_mentions_network}

Using the @-mentions in the tweets we collected, we extracted an evolving social network to use for our investigation. This process was rather involved, for two reasons:

\begin{enumerate}
\item Because the snowball sampling data collection process itself took several weeks, and because we collected only the last 200 tweets for each user, the time period for which we had data
was not the same for all users. Thus we needed to balance the desire for an evolving network covering a longer period with the desire to have complete data for as many users as possible for that time period.

\item We wanted to focus our analysis on ordinary human users of Twitter, so we wanted to screen out outlier users such as celebrities and bots. Celebrity accounts tend to be mentioned
by a vast number of users, and some types of bot mechanically mention huge numbers of users. Including these accounts could cause the network structure to become
degenerate, with a path of length two existing between most pairs of users via an intermediate celebrity or bot.
\end{enumerate}

We extracted an evolving mentions network for the seven-day period from 9th October to 15th October 2014, consisting of 6,052,615 edges between
285,168 users. These edges came from 4,389,362 tweets (one tweet can mention multiple users, giving rise to more than one edge).
Details of the extraction and filtering steps are given in Appendix~\ref{app:extracting_network}.
We calculated a broadcast scores for each user, using a range of values of $\alpha$: 0.15, 0.3, 0.45, 0.6, 0.75 and 0.9.

\begin{figure}[t]
\centering
\includegraphics[width=\textwidth]{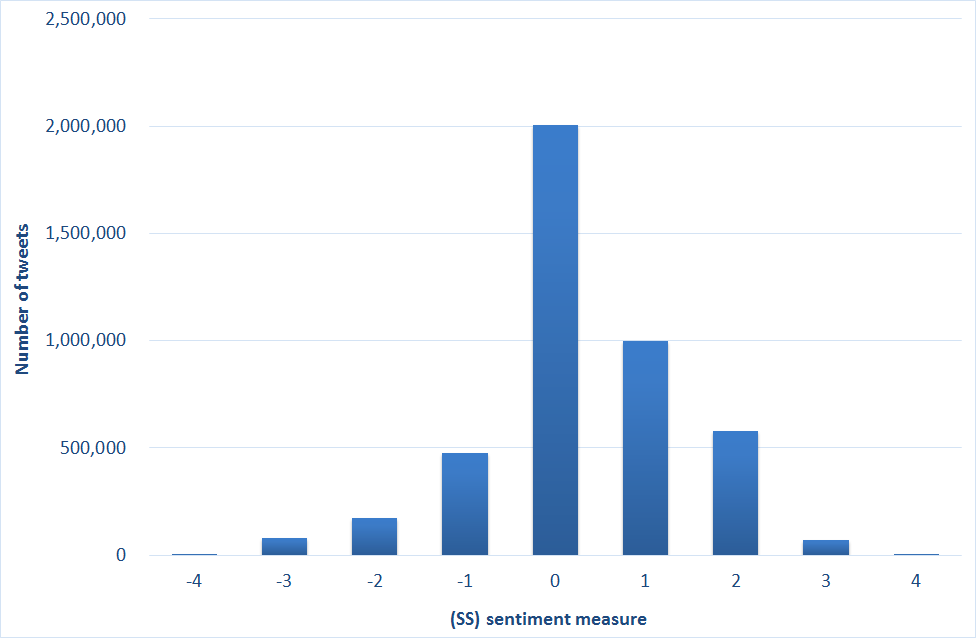}
\caption{Histogram of the (SS) sentiment measure scores for the tweets in the mentions network we analysed.}
\label{fig:hist_sent}
\end{figure}

The distribution of the (SS) scores for all the tweets in our one-week network is shown in Fig.~\ref{fig:hist_sent}. The mean sentiment was mildly positive for all three measures:
0.297 for (SS), 0.823 for (MC) and 3.669 for (L).
The limitations of the sentiment scoring algorithms explain the high proportion of tweets assgined a zero score (as shown for example in Fig.~\ref{fig:hist_sent}). Some of these are genuinely tweets with a neutral tone, but some are tweets where the algorithm cannot detect any sentiment, so we think of the zero score as indicating ``neutral or not detected'' sentiment.
At the level of individual tweets, the Pearson correlation coefficients between the three sentiment measures (MC), (SS) and (L) are as follows:

\begin{center}
\begin{tabular}{l l}
 (MC) and (SS): & 0.585 \\
 (MC) and (L): & 0.524 \\
 (SS) and (L): & 0.564
\end{tabular}
\end{center}
Although the correlations at the individual tweet level are moderate, we will later see in Section~4.\ref{sub:endurance} that when we aggregate to groups of tweets, such as all the tweets sent within
a particular community, the correlations become very strong.

\subsection{Broadcast scores vs. average sentiment}

\begin{figure}[ht]
\centering
\includegraphics[width=\textwidth]{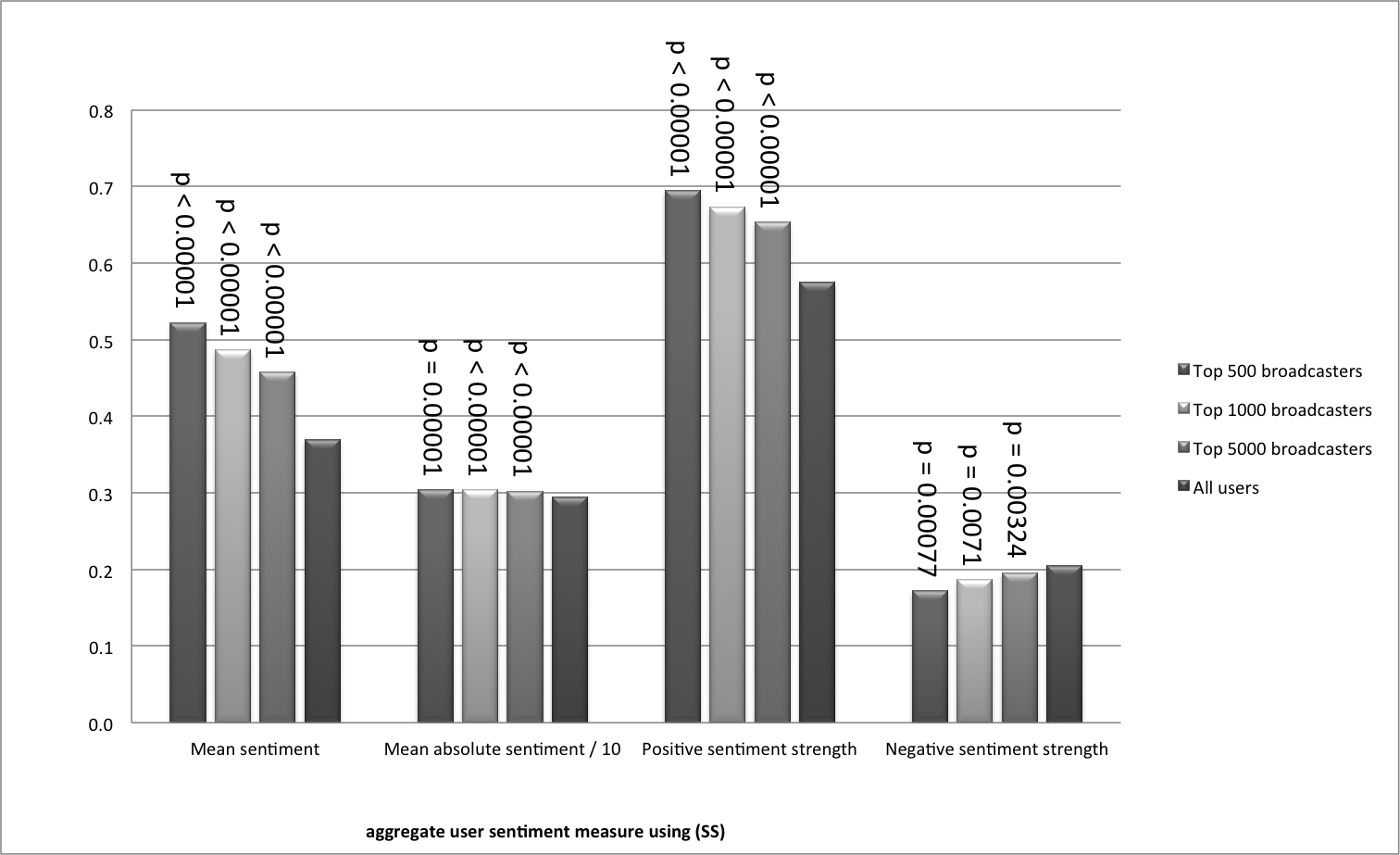}
\caption{The means of the (SS) sentiment attributes for the top 500, 1000 and 5000 broadcasters (for $\alpha = 0.75$) compared with the mean values across all users. (The mean absolute sentiment values have been divided by 10 for easier viewing.)}
\label{fig:sent_top_broadcasters}
\end{figure}

\begin{figure}[ht]
\centering
\includegraphics[width=\textwidth]{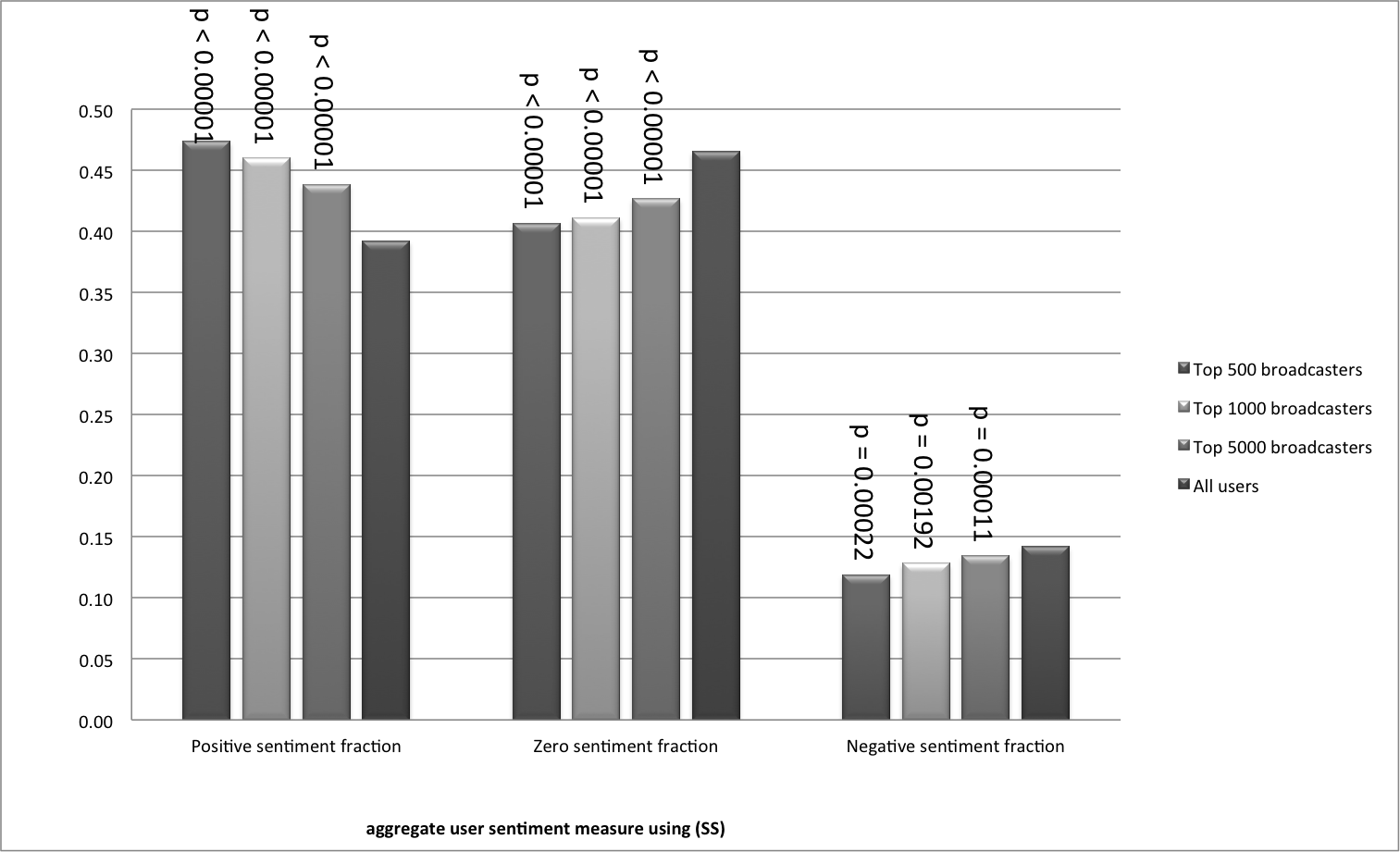}
\caption{The means of the (SS) sentiment fraction attributes for the top 500, 1000 and 5000 broadcasters (for $\alpha = 0.75$) compared with the mean values across all users.}
\label{fig:sent_frac_top_broadcasters}
\end{figure}

We now compare broadcast scores with users' sentiment use. For this we need user-level sentiment attributes, but the three sentiment scoring algorithms that we used assign a sentiment score to each tweet. Therefore we aggregated the sentiment scores of each user's outgoing edges within the network, to get the following seven attributes (for each of the three measures):

\begin{itemize}
\item	Mean sentiment: the mean of the sentiment scores for the user's outgoing edges.
\item	Mean absolute sentiment: the mean of the absolute values of the sentiment scores for the user's outgoing edges.
\item	Positive sentiment fraction: the fraction of the user's outgoing edges having a sentiment score greater than zero.
\item	Zero sentiment fraction: The fraction of the user's outgoing edges having a zero sentiment score (indicating a neutral sentiment or that no sentiment could be identified by the scoring system).
\item	Negative sentiment fraction: the fraction of the user's outgoing edges having a sentiment score less than zero.
\item	Average positive sentiment strength: the sum of the user's sentiment scores over the outgoing edges with positive scores only, divided by the count of the user's outgoing edges (this count includes all outgoing edges the user sent, not just those with a positive score).
\item	Average negative sentiment strength: the sum of the absolute values of the user's sentiment scores over the outgoing edges with negative scores only, divided by the count of the user's outgoing edges (this count includes all outgoing edges the user sent, not just those with a negative score).
\end{itemize}
The purpose of the two sentiment strength attributes is to take into account not just how often a user expresses positive or negative sentiment, but also how extreme that sentiment is when it is expressed.

Users with no outgoing edges on the first day of our studied seven-day evolving network are at a disadvantage in terms of broadcast scores, because their messages have only six (or fewer) days to propagate through the network, rather than seven. So for the rest of this section we report on just the 153,691 users who tweeted within the network on the first day.

In Fig.~\ref{fig:sent_top_broadcasters} and Fig.~\ref{fig:sent_frac_top_broadcasters} we compare the means of the above attributes for the top 500, 1000 and 5000 broadcasters with the means over all users, using (SS) and $\alpha=0.75$. We see that:
\begin{itemize}
 \item Top broadcasters send messages with positive sentiment more frequently, and neutral and negative sentiment less often.
 \item When we additionally account for the extremity of the sentiment that is used as well as the frequency, top broadcasters use more positive sentiment, and less neutral and negative sentiment.
\end{itemize}
The differences are most pronounced for the top 500 broadcasters; as we move from the top 500 to the top 1000 and then top 5000, the means for the top broadcasters gradually become closer to the means for the whole population of users. But even for the top 5000 broadcasters there are still substantial differences.
To confirm the statistical significance of this finding, we have used randomisation testing to estimate (one-sided) p-values\footnote{To explain how these are produced, we shall sketch the calculation of the p-value for one of the attributes, the negative sentiment fraction as shown in Fig.~\ref{fig:sent_frac_top_broadcasters}. The average across all users is 0.142, whereas for the top 500 broadcasters it is only 0.119. We randomly generated 100,000 subsets of the 153,691 users and calculated the means for those subsets. From this we estimate how the mean of the attribute is distributed for randomly chosen sets of size 500. From this distribution we calculate the p-value as the probability that a randomly selected set of 500 users would have a mean equal to 0.119 or more extreme (smaller). This probability is very close to zero (0.00022). Informally, this means we can be very confident that the relationship we have found --- that the top broadcasters use negative sentiment less often --- has not simply happened ``by chance''; the odds of that are less than 3 in 10,000.} which are shown as annotations in Fig.~\ref{fig:sent_top_broadcasters} and Fig.~\ref{fig:sent_frac_top_broadcasters}.

\begin{figure}[t]
\centering
\includegraphics[width=\textwidth]{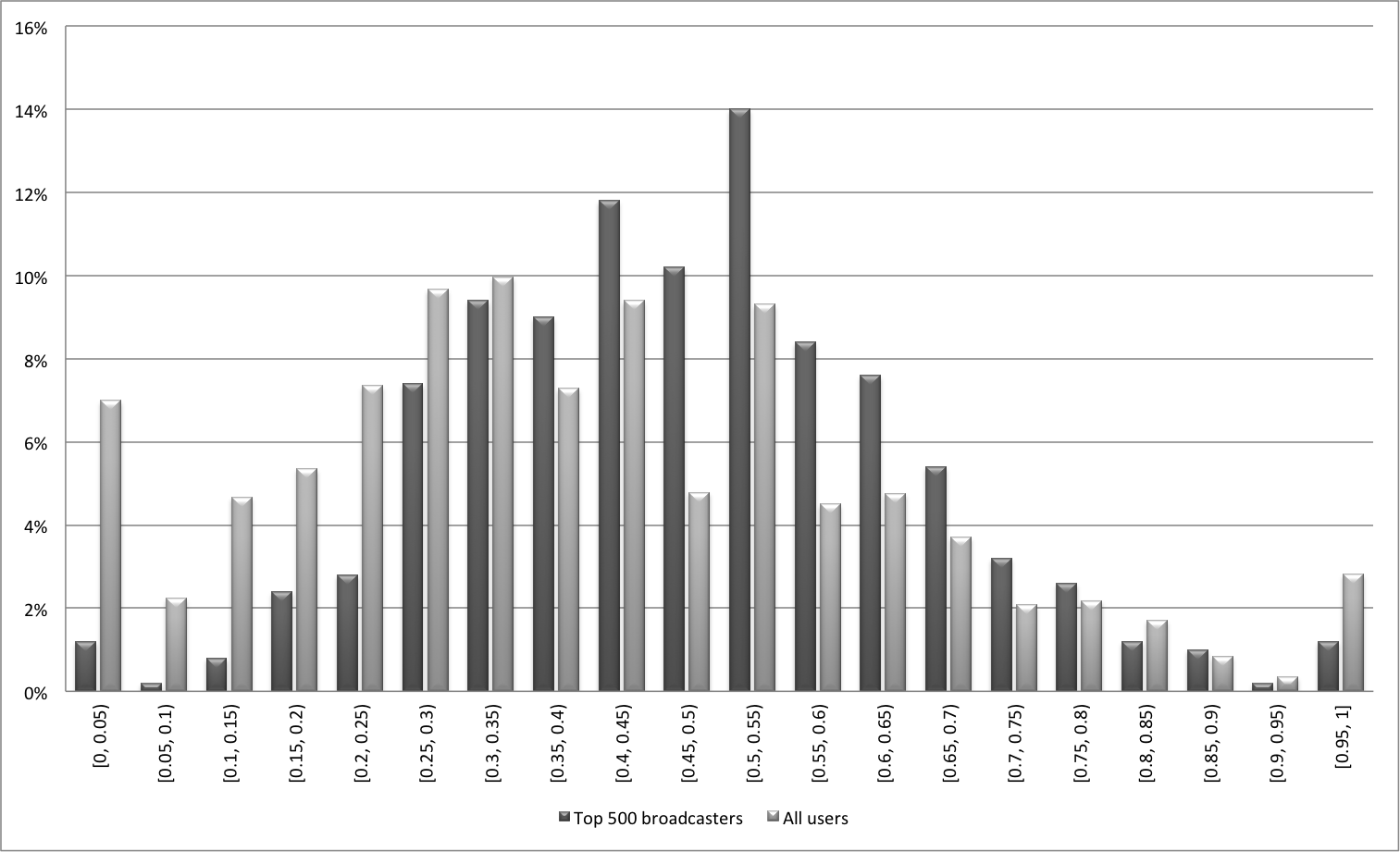}
\caption{Distribution of positive sentiment fraction for top the 500 broadcasters (for $\alpha = 0.75$), and for all users, using (SS).}
\label{fig:skewed_dist}
\end{figure}

Note that this does not mean that every user in the top 500 has a higher positive sentiment fraction (i.e.\ uses positive sentiment more frequently) than the average user. Fig.~\ref{fig:skewed_dist} shows the distribution of positive sentiment fraction for the top 500 broadcasters, and for all users, using (SS). The distributions overlap, of course; in particular there are a few top broadcasters with low positive sentiment fractions. Nevertheless, one can clearly see that the distributions are not the same: the distribution for the top 500 broadcasters is in general shifted towards the higher end of the horizontal axis, showing that on average top broadcasters use positive sentiment more often.

Although we have shown the results for (SS) and $\alpha=0.75$, with one exception the same pattern of results was found for all tested values of $\alpha$, and also using the other
sentiment measures (MC) and (L) (again for six tested values of $\alpha$), and the p-values were all less than $0.026$. The exception was that for (SS) and $\alpha=0.15$, the ``negative sentiment strength'' and ``negative sentiment fraction'' attributes for the top 5000 broadcasters were very nearly equal to the mean over all users.

\begin{figure}[t]
\centering
\includegraphics[width=\textwidth]{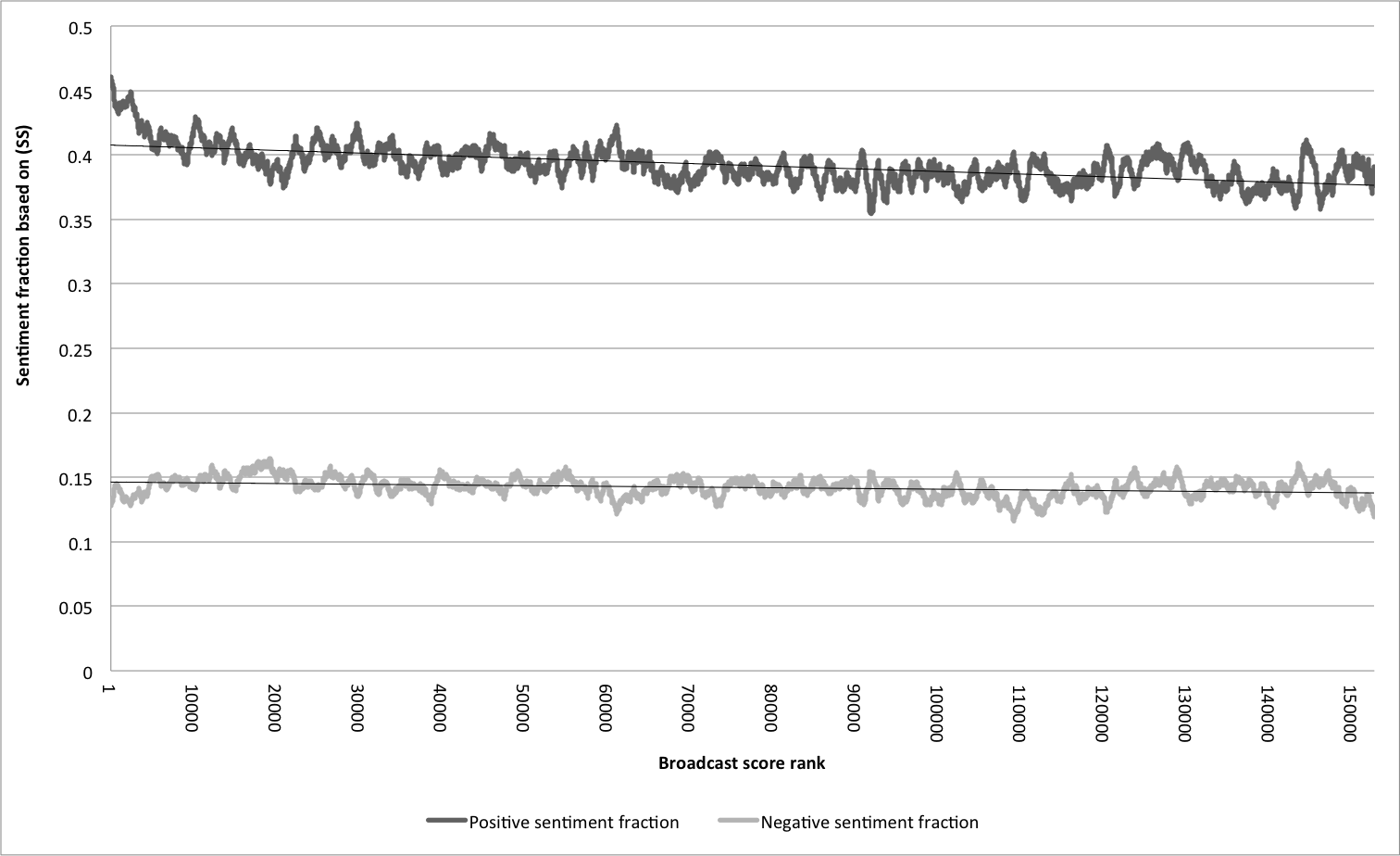}
\caption{The relationship between sentiment fractions (as a moving average over a window of 1000 points) and broadcast score rank, for $\alpha = 0.75$ based on (SS).}
\label{fig:moving_avg_frac}
\end{figure}

\begin{figure}[t]
\centering
\includegraphics[width=\textwidth]{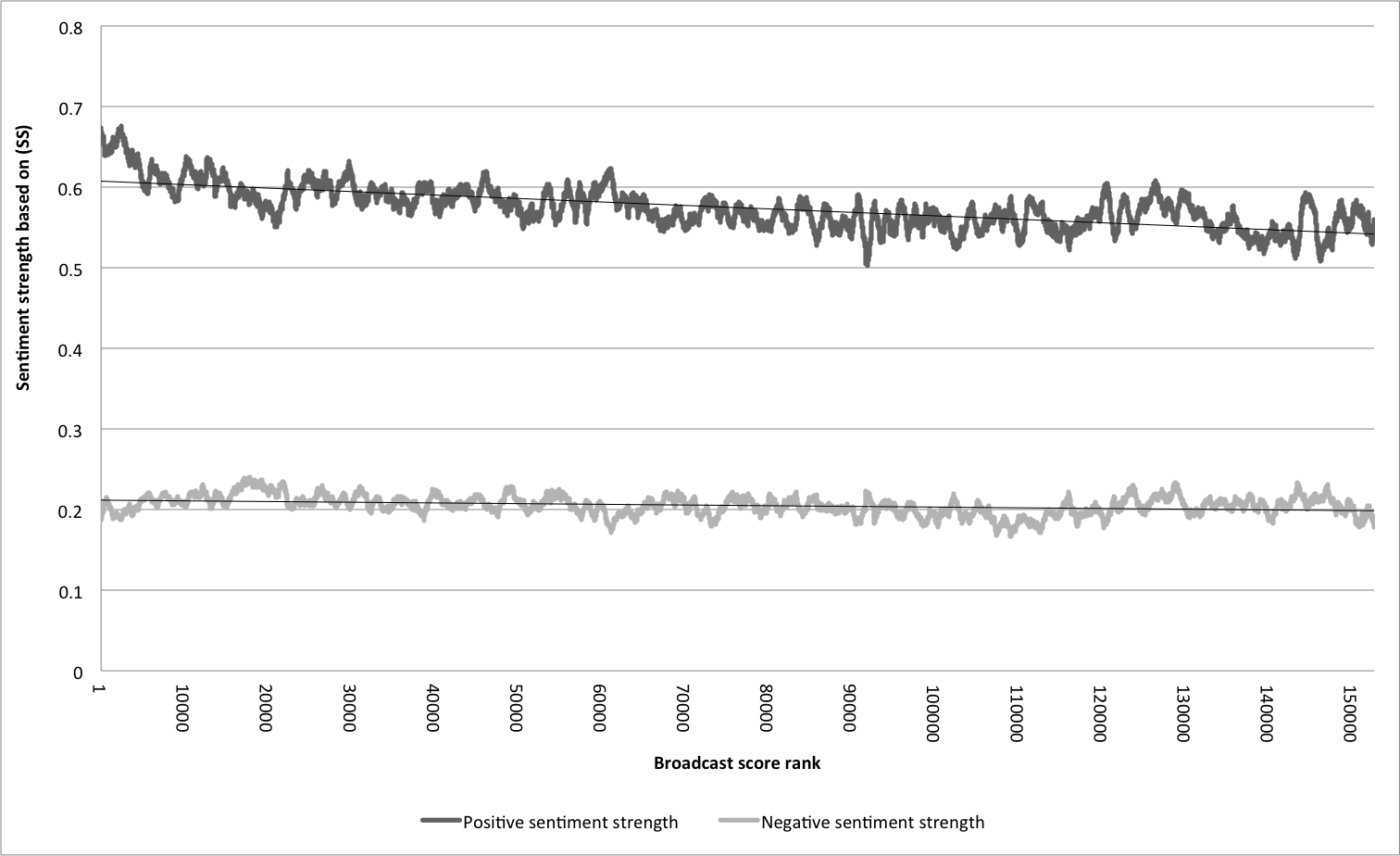}
\caption{The relationship between sentiment strengths (as a moving average over a window of 1000 points) and broadcast score rank, for $\alpha = 0.75$ based on (SS).}
\label{fig:moving_avg_strength}
\end{figure}

In addition to investigating the sentiment use of the top broadcasters, we looked for general trends relating sentiment use to broadcast rank.
Fig.~\ref{fig:moving_avg_frac} plots moving averages of the (SS) sentiment fraction attributes against broadcast rank, using a window of 1000 observations to smooth the noisy data. We see that from rank 1 to about rank 9000 the positive sentiment fraction decreases sharply; after this it decreases slowly in an approximately linear way. The fraction of tweets with negative sentiment appears approximately constant at this scale.
Fig.~\ref{fig:moving_avg_strength} plots similar moving averages for the sentiment strength attributes. The average strength of positive sentiment declines sharply to begin with and then slowly, whereas the average strength of negative sentiment is approximately constant. Although the local fluctuations were different, the graphs had the same general shape for all the values of $\alpha \in \{0.3, 0.45, 0.6, 0.75, 0.9\}$ tested.

\section{Sentiment and evolution of communities on Twitter}

In this section we describe how we identified meaningful communites or ``sub-networks'' of Twitter users, and we present the results of our analysis of how these communities evolved over time, including how their sentiment evolved.

The existence of communities has been observed in all kinds of real-world networks and identifying them has been the subject of considerable research effort in recent years, much of which can be traced back to a seminal paper of Girvan and Newman \cite{Girvan:2002}.
In the vast literature on community detection (see e.g.\ the review article \cite{Fortunato:2010}), a community is often taken to be a group of users with two characteristics:
\begin{enumerate}
\item The community is densely connected internally, i.e. people within the same community talk to each other a lot.
\item There are relatively few links crossing from the community to the outside world, i.e. people talk to fellow members of their community more often than they talk to non-members.
\end{enumerate}

\subsection{How we detected communities and selected a subset for further study}
\label{sub:community_detection}

Because we wanted to find communities that would endure over time, we needed to take a longer period of data than the seven days we analysed in Section~\ref{sec:communicability_and_sent}.
We can imagine online discussions that spring up, rage feverishly for a few days and then largely disappear\footnote{We suspect, for example, that the much-reported discussion of ``What colour is the dress?'' fits into this category; see http://www.bbc.co.uk/news/blogs-trending-31659395.}, and that is not what we wanted to find. Yet, as described in Section~\ref{sec:data}, we had only the last 200 tweets per user, so we needed to limit ourselves to a period where the data was most complete.
We extracted a mentions network from 22nd Sept 2014 (inclusive) until the end of our snowball-sampled data, 6th November 2014, a period of 46 days.
The process for creating the network was the same as described for the seven-day network, described in Appendix~\ref{app:extracting_network}.
The resulting network consisted of 491,417 users with 31,299,836 edges between them, coming from 22,594,048 tweets. For the first 40 days the daily average was 776k edges; for the last 6 days, when data collection was coming to an end, the daily average was only 40k edges.
The network has an average of 63.7 outgoing edges per user, corresponding to 46.0 tweets per user, and each user mentioned an average of 30.9 distinct recipients.

With the dataset chosen, we turn to the question of algorithms. Discovering communities by algorithms requires one to first formulate a precise definition of how ``good'' a given division of a social network into communities is. The most widely used formula for quantifying the ``goodness'' of a division is called modularity \cite{Newman:2004} and it compares the fraction of edges that lie within a community in the network with the expected fraction of edges that would lie within the community if the edges were placed at random. Many different versions of modularity have been proposed in the last decade. As we look at relatively unbalanced divisions (trying to identify small portions of a large network), we considered instead a different measure called conductance \cite{Brandes:2005} which takes values from $0$ to $1$. Groups of users that are well connected internally but well separated from the rest of the network have values close to $0$, and groups with few internal connections but lots of connections to the rest of the network have values close to $1$.

There is also a variant of conductance, called weighted conductance, that takes into account the weights on edges, rather than just their presence or absence. We use the number of messages exchanged between two users (in either direction) as the weight of the edge between them. Thus weighted conductance depends not only on which users have corresponded with which others, but also on how often. If $W_{ij}$ is the weight of the edge from user $i$ to user $j$, $S$ is a community, and $\bar{S}$  denotes the remaining users, the weighted conductance of $S$  is \[\frac{\sum_{i \in S, j \in \bar{S}}{W_{ij}}}{\min{(a(S), a(\bar{S})})}\]
where $a(S)=\sum_{i\in S}\sum_{j \in \bar{S}}W_{ij}.$

We used the following three algorithms to identify communities:
\begin{itemize}
\item The Louvain method on unweighted graphs, described in \cite{Blondel:2008}, as implemented in Python
in the library \cite{comms:2009} and in C++ by Lefebvre and Guillaume\footnote{ This code is freely available from https://sites.google.com/site/findcommunities/.}.
\item The Louvain method on weighted graphs, using the C++ implementation.
\item The k-clique-communities method\footnote{This algorithm, and some refinements to it, are also implemented in the CFinder program, freely available from http://www.cfinder.org/.} presented in \cite{Palla:2005} as implemented in the NetworkX Python library.
\end{itemize}
Using these three methods with different parameters, we produced a list of 98,078 candidate communities.
For each community we calculated:
\begin{itemize}
\item the size of the community (number of nodes),
\item the number of internal edges (mentions between users),
\item the number of internal edges (mentions) per node (this gives a measure of how much activity there is inside the community),
\item the conductance  and the weighted conductance of the community within the whole network,
\item the mean sentiment of edges within the community, using the (MC) measure\footnote{Due to time constraints and the large number of tweets involved in community detection, we decided not to calculate
the (SS) and (L) scores at this stage.},
\item whether the community consisted of a single connected component (good candidate communities will of course be connected; however, very infrequently the Louvain method can generate disconnected communities, by removing a ``bridge'' node during its iterative refinement of its communities),
\item the fraction of internal mentions with non-zero sentiment (some of our candidate communities were composed mainly of users speaking a non-English language, and we used this measure to filter them out; tweets in other languages are likely to be assigned a zero sentiment score, because the sentiment scoring algorithm does not find any English words with which to gauge the sentiment),
\item some statistics summarising the role played in the community by recently registered users, and
\item a breakdown of the frequency of participation of users in the community. (For each user in the community, we counted how many distinct days they had been active on Twitter in our data, and then calculated the percentage of these days on which they had posted within the candidate community. We calculated the average across all users in the community, and also split the users up into five bins.)
\end{itemize}
Based on the above statistics we short-listed a subset of communities and performed a manual inspection of a sample of the tweets within the community, to assess the topics talked about, and a visualisation of the community, using the program Visone\footnote{http://visone.info/html/about.html} for this subset.

In the end we selected 18 communities to monitor and study. Table~\ref{comms-stats} shows most of the statistics listed above for these 18 communities, in size order. In each numerical column, the highest six values are highlighted in green and the lowest six values are highlighted in red (recall that for conductance and weighted conductance, lower values indicate a more tightly-knit community). The ``Algorithm'' column contains `L' for the Louvain method, `W' for the weighted Louvain method and `K' for the k-clique-communities method. We chose six communities from each algorithm.

\begin{table}
\centering
\includegraphics[angle=90,height=0.9\textheight]{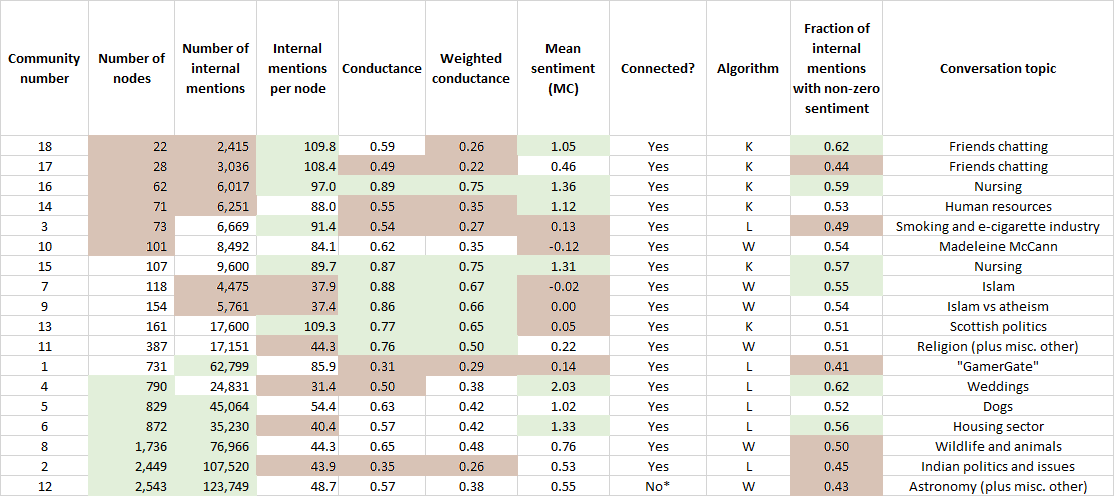}
\caption{Selected summary statistics for the 18 communities we selected, in size order. The community marked as not connected had 8 nodes separated from the rest.}
\label{comms-stats}
\end{table}

Table~\ref{comms-part} shows frequency of participation, with communities ranked by the third column, which gives the average user participation. This is expressed as a percentage: the percentage of days on which the user was active on Twitter (in our dataset) that they were active in the community. The rightmost five columns show, for each community, how the users' participation levels break down into five bins. Bins with disproportionately many users in them (i.e.\ with values more than 0.2) are highlighted in green. We can see that with the exception of community 4 (Weddings), every community has at least a 20\% ``hard core'' of users, who are active in the community nearly every day they are active on Twitter.

Once we had selected the communities of interest, we collected a more detailed tweet history for each participating user, as described in Section~\ref{sec:data}.

\begin{table}
\centering
\includegraphics[width=\textwidth]{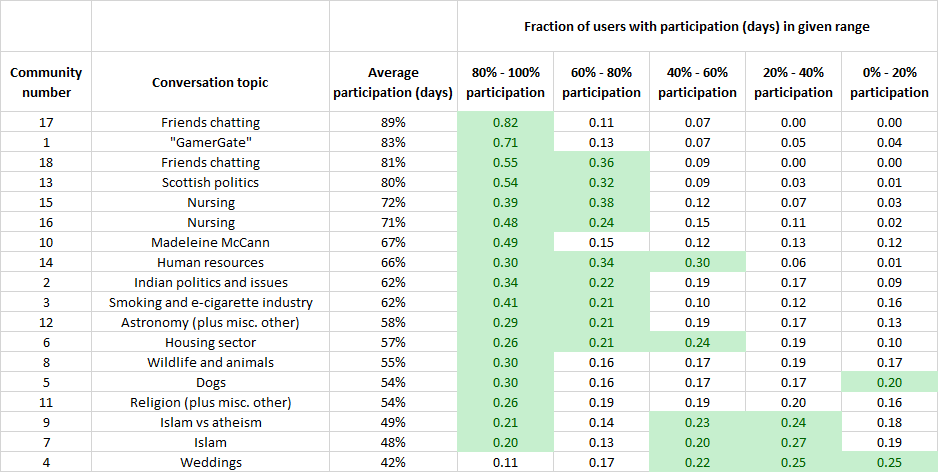}
\caption{Figures for frequency of participation, with communities ranked by average user participation (the third column).}
\label{comms-part}
\end{table}

\subsection{Analysing the endurance of the communities}
\label{sub:endurance}

\begin{figure}[t]
\centering
\includegraphics[width=\textwidth]{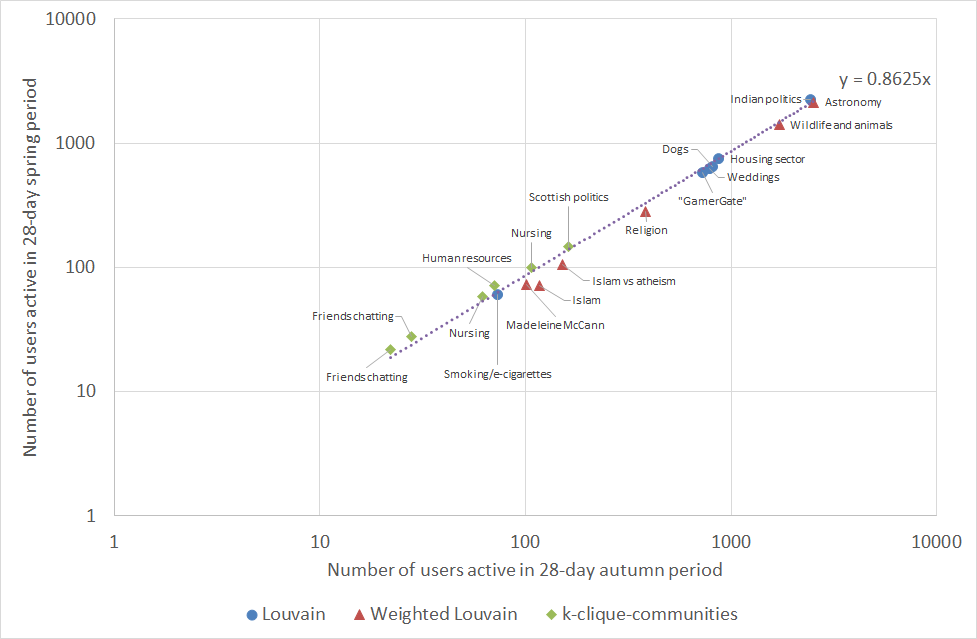}
\caption{The communities we studied endured strongly over a 19-week period.}
\label{fig:endurance_log_log}
\end{figure}

We analysed was how well our communities endured over time.
We examined a 28-day period starting on 22nd September 2014 (which we will call the ``autumn period'') and a 28-day period starting on 2nd February 2015 (which we will call the ``spring period''), and compared how many users in each community were active (mentioned or were mentioned by other users) within the community. Would the same users still be tweeting each other in the spring, or would the communities have dissolved over time? Fig.~\ref{fig:endurance_log_log} shows a log-log plot of the results.

\begin{figure}[h]
\centering
\includegraphics[width=\textwidth]{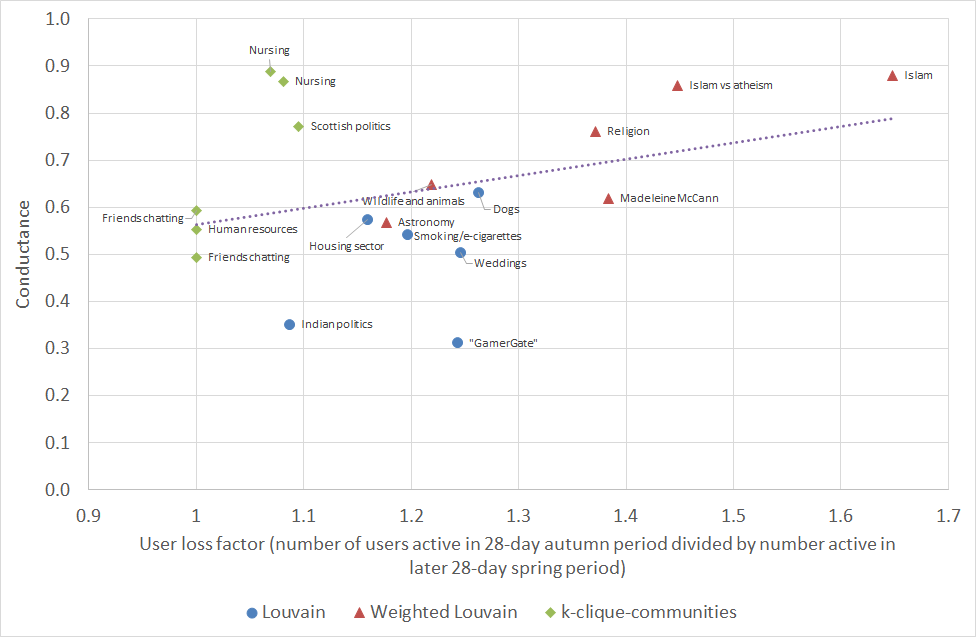}
\caption{Communities with higher conductance tended to lose more of their users over time.}
\label{fig:loss_vs_cond}
\end{figure}

\begin{figure}[h]
\centering
\includegraphics[width=\textwidth]{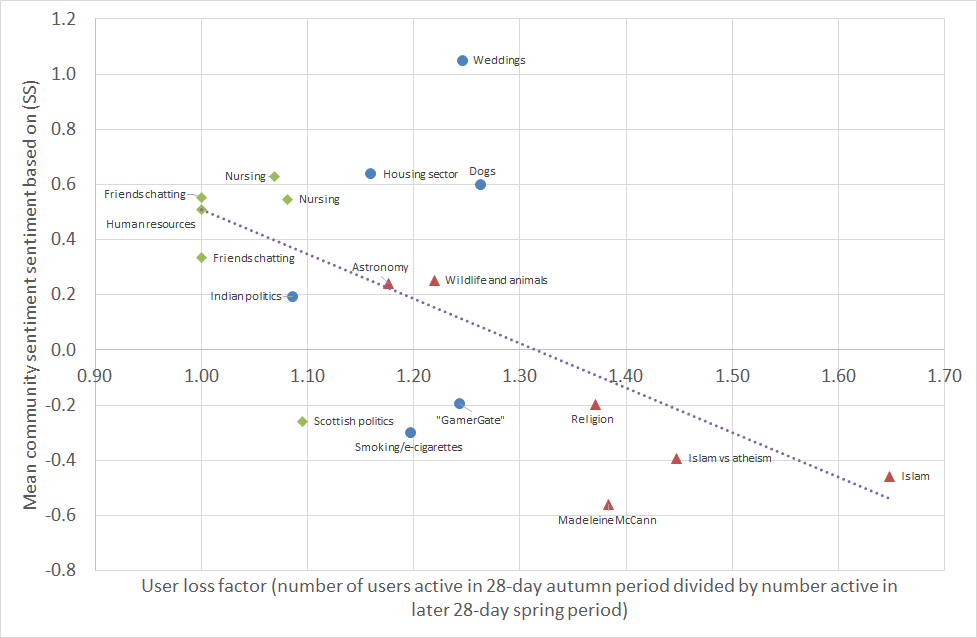}
\caption{Communities with more negative sentiment, measured by (SS), tended to lose more of their users over time.}
\label{fig:loss_vs_sent}
\end{figure}

We see that the communities persisted well from autumn to spring. In three of them, communities 14 (Human resources), 17 (Friends chatting) and 18 (Friends chatting), all the original users were still active in the community. These are three out of the four smallest communities. The other 15 communities lost between 6.5\% (for community 16, Nursing) and 39.3\% (for community 7, Islam) of their users, with an average loss of 18.6\%.
We can see differences in the communities produced by the three algorithms here: the six produced by k-clique-communities lost an average of 3.8\% of their users, compared to 16.4\% for the Louvain method and 26.3\% for weighted Louvain.

Let us say \emph{user loss factor} to mean the number of users active in the 28-day autumn period divided by number active in the later 28-day spring period. When the user loss factor is 1, then the community has retained all its users; the higher the value, the more users the community has lost.
We looked to see whether the conductance, sentiment or size of communities is related to their endurance.
In Fig.~\ref{fig:loss_vs_cond} one can see that conductance is a predictor of what proportion of users will stop participating in the community, with correlation coefficient $0.42$.
When conductance is lower (so that the community is more densely connected internally and better separated from the rest of the network) then fewer users stopped participating on average.

Similarly, the community sentiment is a predictor of community endurance, as shown in Fig.~\ref{fig:loss_vs_sent}: the more positive the initial sentiment (measured in the autumn period), the fewer users stopped participating on average. For (SS) (as shown in Fig.~\ref{fig:loss_vs_sent}) the correlation coefficient is $-0.60$; for (MC) it is $-0.48$ and for (L) it is $-0.58$.
On the other hand community size was not correlated to user loss factor; the correlation coefficient was $0.07$.

We noted in Section~3.\ref{sub:extracting_mentions_network} that the correlations between the three sentiment measures (MC), (SS) and (L) at the individual tweet level were only moderate. The following shows the correlations between the \emph{community} sentiments produced by the three measures, in the autumn and spring periods:

\begin{center}
\begin{tabular}{l c c}
Measures & Correlation coefficient (autumn) & Correlation coefficient (spring)\\
(MC) and (SS) & 0.971 & 0.954\\
(MC) and (L) & 0.972 & 0.929\\
(SS) and (L) & 0.985 & 0.948
\end{tabular}
\end{center}
Thus at the community level the three measures are very similar.

\subsection{Dynamics of sentiments in communities}
\label{sub:dynamics_of_sentiment}

\begin{figure}[t]
\centering
\includegraphics[width=\textwidth]{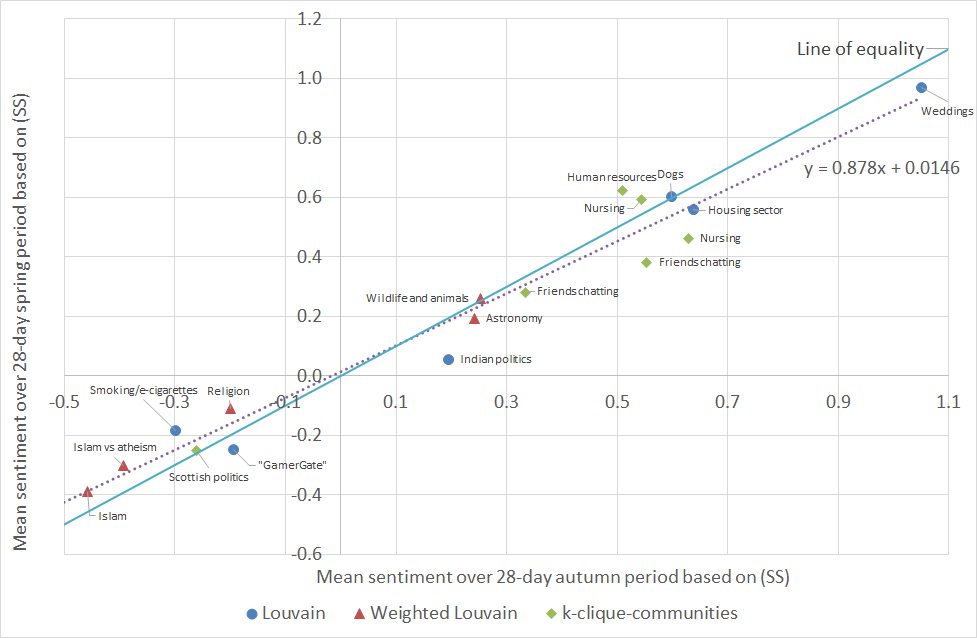}
\caption{Graph showing that community sentiment was very stable over the 19-week period. The solid line shows where the autumn and spring sentiments are equal.}
\label{fig:sent_stability}
\end{figure}

Here, we analyse the changes in sentiment/mood of our communities over time (or the lack thereof, as it generally turns out).
Fig.~\ref{fig:sent_stability} plots the mean (SS) sentiment of each community over the autumn period against the mean (SS) sentiment over the spring period. We see that the sentiments persisted very strongly: the correlation between the autumn sentiment and spring sentiment is $0.982$.
The corresponding correlation under the (MC) measure was $0.982$, and under (L) was $0.960$.

\begin{figure}[t]
\centering
\includegraphics[width=\textwidth]{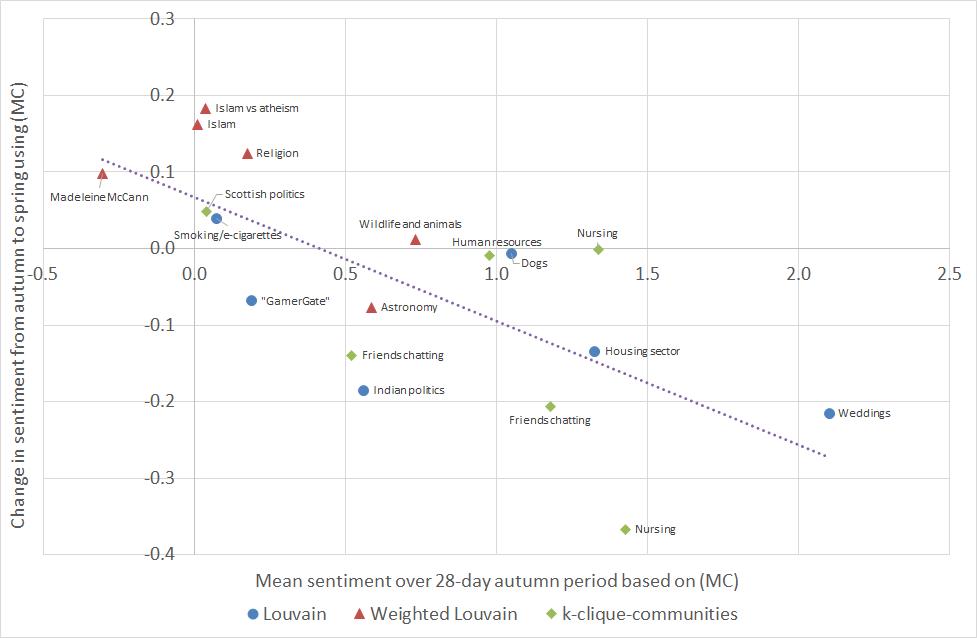}
\caption{Graph showing that the sentiment in 16 of the 18 communities became more moderate over time. This plot uses the (MC) measure.}
\label{fig:sent_changes}
\end{figure}

We looked for explanations for the (small) changes in sentiments that did occur. On the vertical axis of Fig.~\ref{fig:sent_changes} we show the change in mean sentiment between the autumn period and spring period using (MC); a positive number means that the sentiment became more positive over time. On the horizontal axis we show the mean sentiment during the autumn period. What we find is that when the sentiment is initially at the negative end of the spectrum, it tends to increase slightly; on the other hand if the sentiment is initially at the positive end, it tends to decrease slightly.

In fact the sentiment in 16 of the 18 communities moved slightly towards a moderate (MC) value of 0.4 (which is approximately where the line of best fit cuts the horizontal axis in Fig~\ref{fig:sent_changes}). This could be because extreme sentiment in a community is ``whipped up'' by external events and then, once those events are over, tends to dissipate naturally with time.

We point out however that there is likely also an element of statistical ``regression to the mean'' occurring. We did not choose our communities at random: we chose five of them because they were among those with the most extreme sentiment in the autumn period\footnote{Namely: 4 (Weddings), 7 (Islam), 9 (Islam vs atheism),
10 (Madeleine McCann) and 11 (Religion).}. This introduces a bias and makes it more likely for the sentiment in these five communities to become more moderate by the spring period (which it does, in all five cases). This bias is unavoidable when one disproportionately selects communities with extreme sentiment for study.
The correlation coefficient Fig.~\ref{fig:sent_changes} is $-0.71$. The relationship was less apparent using the other sentiment measures, though still present, with corresponding correlations of $-0.59$ for (SS) and $-0.32$ for (L).

The robustness of the weekly sentiment measures suggests that only a limited amount of data, say for two or three weeks, is needed to give a good idea of the sentiment of a Twitter community, and if a drastic change in sentiment does occur within a community, this is a rare event and may indicate that something important has happened to or within the community.

\begin{figure}
\centering
\includegraphics[width=\textwidth]{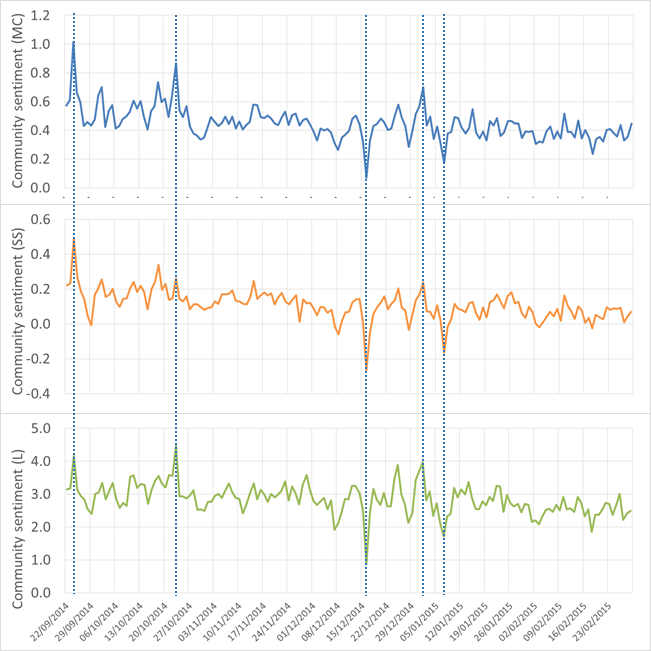}
\caption{The daily mean sentiment in community 2 (Indian politics) from 22nd September 2014 to 1st March 2015. Five interesting events are identified.}
\label{fig:sent_of_indians}
\end{figure}

Looking at the daily average sentiment in each community, that is, looking at a higher resolution, more detail is evident. Fig.~\ref{fig:sent_of_indians} shows the daily mean sentiment in community 2 (Indian politics), also for the period 22nd September 2014 to 1st March 2015.
Large day-to-day variations can be seen, and we have noticed that often such abrupt changes can be traced to real events affecting the community.
In Fig.~\ref{fig:sent_of_indians} we have highlighted five dates where the sentiment measures show spikes or troughs. By examining the tweets sent on those
dates we identified the significant event that drove the sentiment change:
\begin{itemize}
\item 24th September 2014: India's Mars Orbiter Mission space probe entered orbit around Mars, and people celebrated.
\item 23rd October 2014: the beginning of the Diwali festival.
\item 16th December 2014: gunmen affiliated with the Tehrik-i-Taliban conducted a terrorist attack in the northwestern Pakistani city of Peshawar.
\item 1st January 2014: New Year's Day.
\item 7th January 2015: gunmen attacked the offices of the French satirical weekly newspaper Charlie Hebdo in Paris.
\end{itemize}

\section{An agent-based model of sentiments dynamics in communities}
\label{sec:abm}

It has been discovered time after time that the collective behaviour of populations of interacting individuals is difficult to understand, challenging to predict and sometimes even seemingly paradoxical.
In order  to  be able to predict  the likely evolution of sentiment within a community  and to explore its dynamics under various change scenarios, such as the departure of particular users or the arrival of a new vocal user, we built an Agent-Based Model (ABM) of our Twitter communities. This  includes modelling the sentiment of individuals in the network, and how sentiment spreads from one user to another.

The agents in the model represent Twitter users, and they are arranged in a static undirected graph; only pairs of agents connected by an edge are able to exchange messages. The simulation proceeds in discrete time steps; the number of these steps per day is a parameter of the model. At each time step the following things happen:
\begin{itemize}
\item Each agent performs an action which consists of sending a burst of messages to all/some/none of its neighbours, influenced by the agent's current state.
\item Each agent evolves into a new state, influenced by the actions of other agents in this step, i.e. influenced by the messages it has received this step.
\end{itemize}
Specifically, an action by an agent consists of: a subset of neighbours who will be messaged at this time step; for each neighbour messaged, the number of messages sent to them at this time step; for each neighbour messaged, a sentiment for the messages sent to them at this time step.
The state of an agent consists of two variables. The first is a real number representing the current sentiment level of the agent, on the same scale as the sentiment
scores used for messages. The second is a record of who sent a message to the agent recently: this is the subset of the agent's neighbours who sent the agent a message at the previous time step; these are candidates for the agent to reply to.
In addition to its evolving state, each agent $A$ has a set of constant characteristics that influence its behaviour but do not evolve:

\begin{enumerate}
\item an {\bf initiation probability $P(init, A)$} which controls the tendency of the agent to initiate new conversations with other users when it has received no messages recently
\item a {\bf reply probability $P(reply, A)$} which controls the tendency of the agent to reply to messages it has received
\item a {\bf propagation probability $P(prop, A)$} which controls the tendency of the agent to propagate messages, that is, to message some other user $B$ after being prompted by a message from a different user $C$ in the previous time step
\item a {\bf baseline sentiment level $S(baseline ,A)$}: this is the sentiment level the agent starts off with, and to which is may reset from time to time (as described below)
\item a {\bf neutral sentiment level $S(neutral, A)$}: when the agent receives messages with sentiment higher than this level, the agent's sentiment will be raised, and when the agent receives messages with sentiment lower than this level, the agent's sentiment will be lowered
\end{enumerate}
The model also has six global parameters: the number of iterations (discrete time steps) per day, the mean number of messages per burst, a \emph{contagion of sentiment factor},
a \emph{sentiment reset probability}, a \emph{sentiment noise level} and a \emph{neighbour frequency threshold}. The details of the global parameters, how the agents decide to send messages and how the agents' sentiments evolve  are given in Appendix~\ref{app:abm_params_and_rules}.

The process of using this model to simulate a real Twitter community is then as follows.
First we construct the graph from the historical data for the community, connecting the users that have exchanged more messages than the neighbour threshold. We set the baseline sentiment $S(baseline, A)$ for each agent $A$ by computing the mean sentiment of messages sent by each user, and we set the neutral sentiment $S(neutral, A)$ of each agent to the mean sentiment of all messages sent in the community. To estimate the initiation probability $P(init, A)$ for each agent $A$, we split the historical data into windows, with length determined by the number of iterations per day. We count the number of opportunities $A$ had to initiate a conversation (i.e.\ how many windows there were in which $A$ received no messages), and also how many times out of these $A$ actually initiated a conversation. The reply and propagate probabilities  $P(reply, A)$ and $P(prop, A)$ are set similarly.

To perform a simulation run of the model, we set all the agents to their initial state, and then we evolve the system for the required number of steps, recording the messages that were sent for later analysis. The initial state of each agent is that the agent has received no messages to consider replying to, and its current sentiment is equal to its baseline sentiment. The required number of steps is the number of days in the real data multiplied by the number of iterations per day  (so that the time period of the simulation matches that of the real data).

\subsection{Calibration}
We now describe how we calibrated our model to our Twitter data. The purpose of the six global parameters is to make our agent-based model ``tuneable'', so that we can fine-tune it to match the behaviour observed in different kinds of online community. Calibrating the model to a particular community means finding the values of the six parameters that maximise the match between the model and the real data, i.e.\ the parameter values that make the simulation runs of the model most closely resemble the real data.
In our case, the specific metrics that we use to compare the simulated data with the real data are: the activity levels (number of messages sent per day) of each individual user, and the day-to-day volatility of this, as well as the sentiment of the whole network, and its day-to-day volatility. Comparing the real data and simulated data in this way is an instance of the \emph{method of simulated moments}.

We therefore propose the following function $\rho$ to score a particular simulation run (smaller scores mean a better match):
\begin{displaymath}
\rho=\alpha \sum_{i=1}^N\left|\langle C_i \rangle -  \langle \hat C_i \rangle\right|
+ \beta\sum_{i=1}^N\left| std(C_i) - std(\hat C_i )\right| +
\gamma\left|\langle E_c\rangle - \langle \hat E_c \rangle\right| +
\delta\left|std(E_c) - std(\hat E_c )\right|
\end{displaymath}
Here $N$ is the number of users.  We denote with $\langle C_i \rangle$, $std(C_i) $ the average and standard deviation respectively of the number of messages  sent each day by user $i$ in real data ,  and with $\langle\hat C_i \rangle$, $std(\hat C_i )$ the corresponding values in the simulation run.  Similarly, $\langle E_c\rangle$, $std(E_c)$ denote the average and standard deviation of daily community sentiment and  $\langle \hat E_c \rangle$, $std(\hat E_c)$ those values in the simulation run. The relative sizes of the constants $\alpha, \beta, \gamma$ and $\delta$ are set to reflect how we prioritise the various aspects of the comparison between the real and simulated data. We have used $\alpha=1, \beta=0.1, \gamma=10$ and $\delta=100$ which means that were are putting a lot of emphasis on matching the volatility of daily community sentiment, and less emphasis on matching the level of daily community sentiment. Conversely for the number of messages sent per day by each agent, we prioritise matching the level over matching the volatility.

We chose to model a small community so that we could trace through the simulations, in order to understand them better. We concentrated on modelling community 17 (friends chatting) which has 28 users.
We calibrated the model for each of the three sentiment measures (MC), (SS) and (L).
Each calibration was performed with an iterative grid search: we used five successive grid searches, each time zooming in on the area of the parameter space that appeared most promising in the previous search. The initial ranges searched for each parameter are given in Appendix~\ref{app:param_space}.
Because the simulation runs are randomised we performed 50 simulation runs for each combination of parameters tested, taking the mean of the resulting 50 scores as the score for the choice of parameters.
The parameters found by the repeated grid search were as follows:
\begin{center}
\begin{tabular}{l c c c}
 & (MC) & (SS) & (L)\\
number of iterations per day & 1,536 & 1,536 & 1,536 \\
mean number of messages per burst & 2.1 & 2.1 & 2.1 \\
contagion of sentiment factor & 0.2 & 0.1 & 0.1 \\
sentiment reset probability & 0.03 & 0.13 & 0.15 \\
sentiment noise level & 1.5 & 1.0 & 10.0 \\
neighbour frequency threshold & 18 & 18 & 18
\end{tabular}
\end{center}

\begin{figure}[t]
\centering
\includegraphics[width=\textwidth]{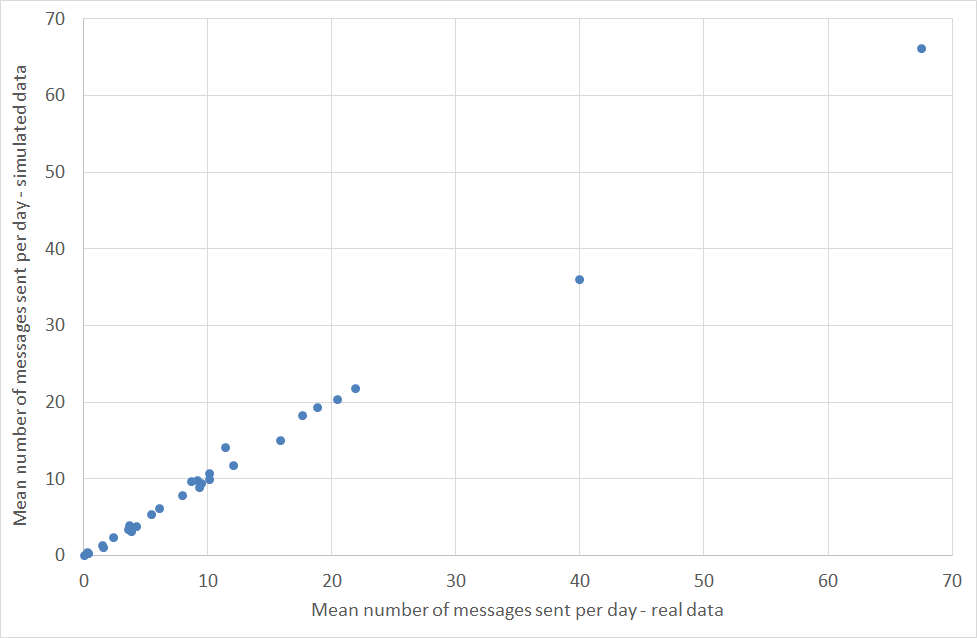}
\caption{The mean daily count of messages sent by each user, in the real data and averaged over 500 simulation runs.}
\label{fig:abm_user_mean}
\end{figure}

\begin{figure}[t]
\centering
\includegraphics[width=\textwidth]{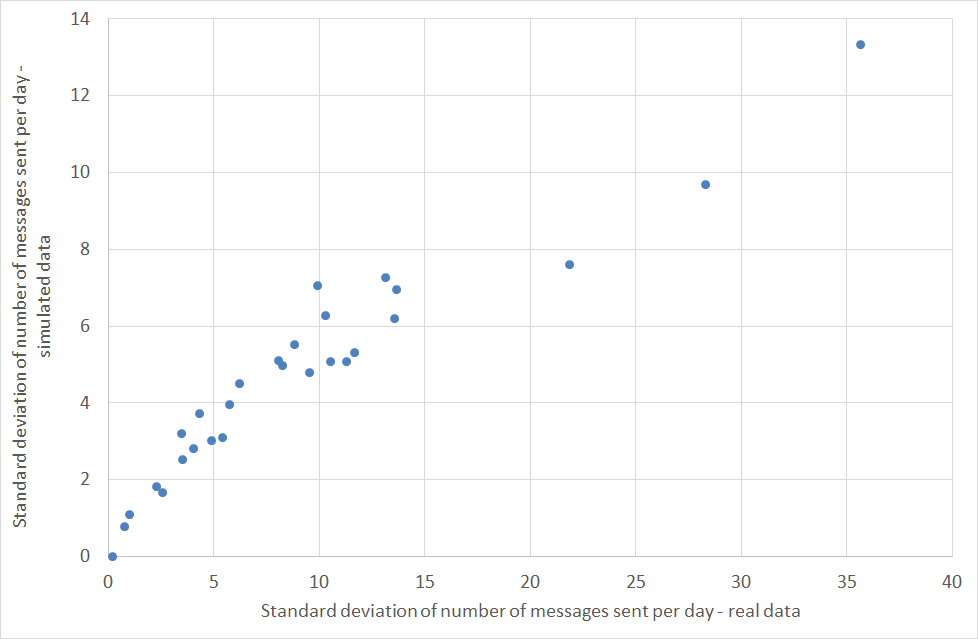}
\caption{The standard deviation (variability) of the daily count of messages sent by each user, in the real data and averaged over 500 simulation runs.}
\label{fig:abm_user_stddev}
\end{figure}

Fig.~\ref{fig:abm_user_mean} compares the mean daily count of messages sent for each user, in the real data and averaged over 500 simulation runs.
As we can see, the match is extremely close. Fig.~\ref{fig:abm_user_stddev} similarly compares the standard deviation (variability) of the daily count of messages sent for each user, in the real data and averaged over 500 simulation runs. The match is less good here, which reflects the fact that when setting the constants $\alpha$ and $\beta$ in the scoring function  we chose to prioritise matching the means instead of the standard deviations.
The sentiment statistics of the real data are matched closely by the simulated data (again averaged over 500 simulation runs), particularly for (MC):
\begin{center}
\begin{tabular}{l c c c}
& (MC) & (SS) & (L)\\
mean daily sentiment in real data & 0.479 & 0.325 & 3.24 \\
mean daily sentiment in simulation & 0.469 & 0.320 & 3.17 \\
standard deviation of daily sentiment in real data & 0.160 & 0.101 & 0.971\\
standard deviation of daily sentiment in simulation & 0.161 & 0.097 & 0.932
\end{tabular}
\end{center}

\begin{figure}[t]
\centering
\includegraphics[width=\textwidth]{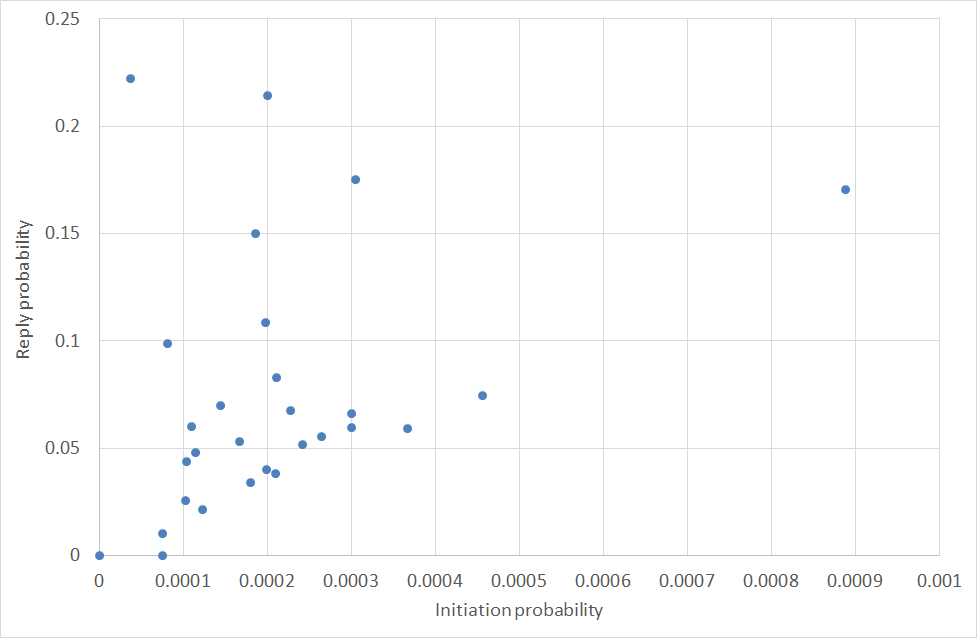}
\caption{Comparing the initiation probability and reply probability for each agent.}
\label{fig:init_vs_reply}
\end{figure}

Finally, in Fig.~\ref{fig:init_vs_reply} we plot the initiation probability $P(init, A)$ of each agent against the reply probability $P(reply, A)$; recall that these are set from the historical data of the community. We include this plot to emphasise the lack of correlation between the two. This confirms that users really do appear to play different roles in the community, with some initiating relatively often but not replying much, and others replying readily while initiating but little.

\subsection{Predicting the effects of introducing a new user}
\begin{figure}[th]
\centering
\includegraphics[width=0.7\textwidth]{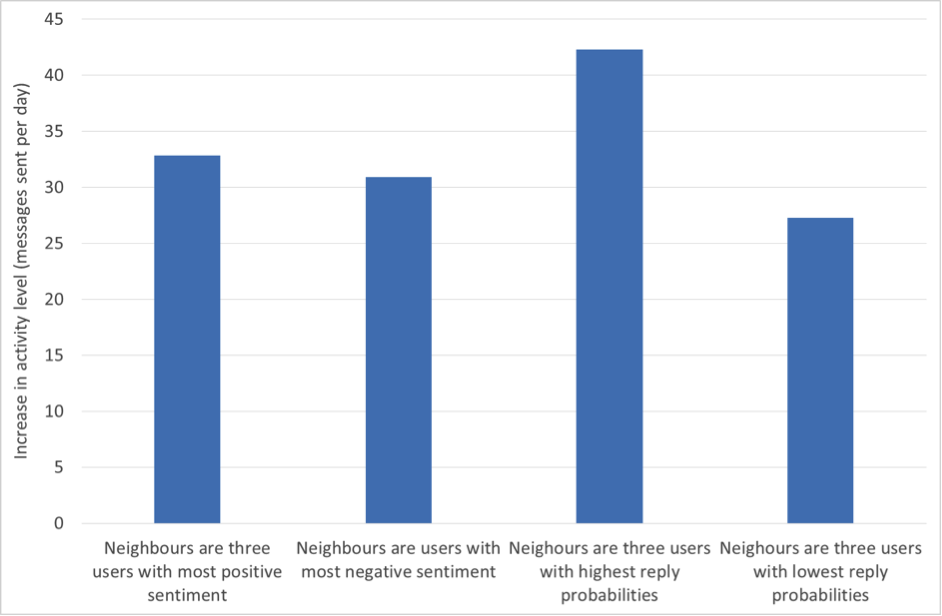}
\caption{The effect on the increase in community activity level of four options for the neighbours of the newly introduced user.}
\label{fig:scenario_activity_lvl}
\end{figure}

\begin{figure}[th]
\centering
\includegraphics[width=0.6\textwidth]{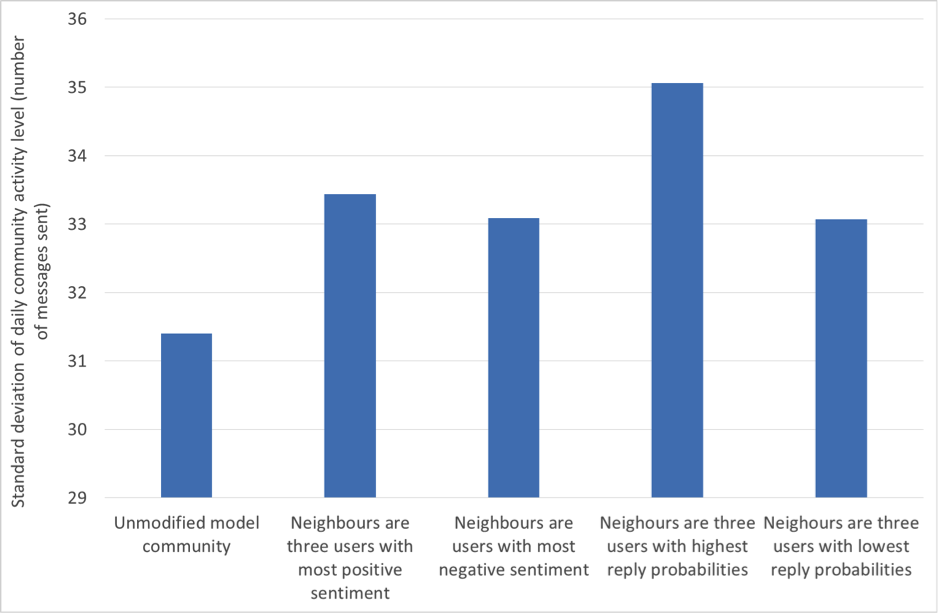}
\caption{The effect on the standard deviation (variability) of daily community activity level of four options for the neighbours of the newly introduced user.}
\label{fig:scenario_activity_variation}
\end{figure}

\begin{figure}[th]
\centering
\includegraphics[width=0.6\textwidth]{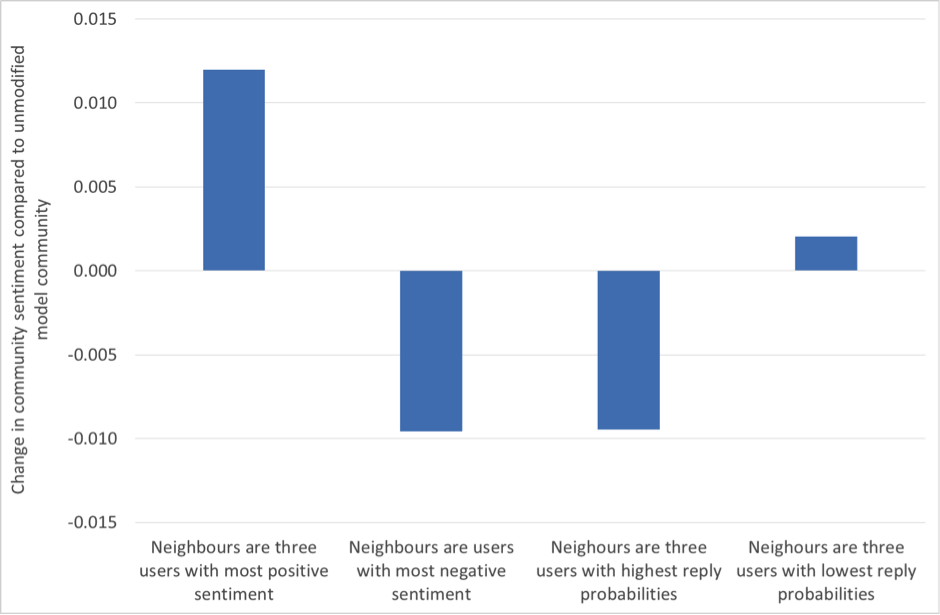}
\caption{The effect on community sentiment level of four options for the neighbours of the newly introduced user.}
\label{fig:scenario_sent_lvl}
\end{figure}

\begin{figure}[th]
\centering
\includegraphics[width=0.6\textwidth]{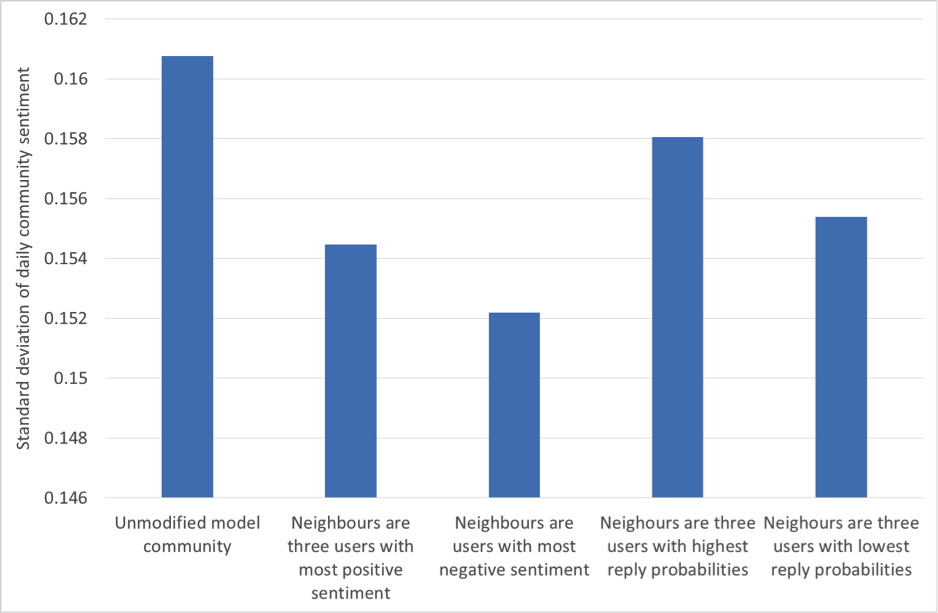}
\caption{The effect on the standard deviation (variability) of daily community sentiment level of four options for the neighbours of the newly introduced user.}
\label{fig:scenario_sent_variation}
\end{figure}

We now consider a scenario where a new user joins the network, and becomes the neighbour of any three existing community members that we choose. Which three community members should our new user befriend? For illustration we explore four possible choices:
\begin{enumerate}
\item	Befriend the three users with the most positive sentiment
\item	Befriend the three users with the most negative sentiment
\item	Befriend the three users with the highest reply probabilities
\item	Befriend the three users with the lowest reply probabilities
\end{enumerate}
For the purposes of this example, we assume that our user will be vocal but with sentiment matched to the prevailing sentiment of the existing community:
The new user's initiation (resp.\ reply, propagation) probability is set to three times the maximum initiation (resp.\ reply, propagation) probability found in the existing community.
Also, the new user's baseline sentiment level is set to the existing community sentiment level.
Figs.~\ref{fig:scenario_activity_lvl}, \ref{fig:scenario_activity_variation}, \ref{fig:scenario_sent_lvl} and \ref{fig:scenario_sent_variation}
 show how our four choices of neighbours affect four aspects of the community: the activity level, the standard deviation (variability) of the daily activity levels, the sentiment level and the standard deviation (variability) of the daily sentiment levels (all averaged over 100 simulation runs, and using (MC)). The results highlight again the role of network structure: if our new user befriends the three most positive users, then the community sentiment goes up, and if he befriends the three most negative users, the community sentiment goes down. Similarly, choosing the users with the highest or lowest reply probabilities as neighbours has a markedly different effect on activity levels.
Validating our model's predictions about the effects of new users on real data is beyond the scope of this paper; it is a challenging research task in itself and is left as future work.

\section{Discussion}
Despite the deluge of data on human communication, dynamics of collective mood is still mainly an uncharted area. While different theories of emotion contagion exist in the literature,  we are still far off being able to predict the occurrences,  intensity and durations of collective compassion, happiness or outrage on Twitter.
Here we presented findings from one large Twitter dataset. While we are conscious of some serious limitations of our approach --- the lack of representativeness of Twitter users, and the noisy nature of sentiment scores --- we believe that our methodology can be generalised to other datasets of human interactions which allow for sentiment scoring.

Looking to wider socio-economic horizons and smart cities opportunities, social media is slowly but steadily becoming an important channel \emph{to run policy information and education campaigns on a mass scale}. Additionally, it has become an exclusive channel to get the attention of some socio-demographic groups, especially in the younger population, who decreasingly consume traditional media such as local newspapers and television.

For these reasons, a data driven model of
collective sentiment captured through social media is one of the most important tools that social data analytics can offer to a city leadership.
It allows gauging public opinion on different topics and understanding/predicting the dynamics of public opinion. Most importantly, it can help to uncover public evaluation of local decisions.
It also allows, as mentioned previously, to engage different communities into a conversation and  to reach to under-represented groups. Our framework can be applied over a wide range of topics: energy, transport, eduction,  tourism, local leadership and so on.

We demonstrated that by using a number of community detection algorithms in combination with sentiment scores, we can identify stable communities of Twitter users. Users within these communities are well-connected and send messages to each other frequently compared with how frequently they send messages to users not in the community. The communities and their ``community sentiment'' were relatively  stable over a time-scale of months. More loose-knit communities and communities with more negative sentiment tended to lose more users over time.  We find that when the sentiment in a community temporarily shows a large deviation from its usual level, this can typically be traced to a significant identifiable event affecting the community, sometimes an external news event.

We have developed an Agent-Based Model of online social networks. The model consists of a population of simulated users, each with its own individual characteristics, such as its tendency to initiate new conversations, its tendency to reply when messaged, and its usual sentiment level. The model allows for sentiment contagion. We have demonstrated that this model, when calibrated with the  data from a real Twitter community, accurately reproduces activity levels and sentiment strength of that community. We have shown an example of using the Agent-Based Model for exploring ``what if\ldots?'' scenarios, such as  ``What if we encourage a new user to interact with particular users in the community?''. To do this we fit the parameters of the model to a particular social network and then make the corresponding modifications to the model. By running a large number of simulations on the modified model, we obtain a prediction of the likely effect of the change on the activity levels and sentiment levels of the community.

\section*{Acknowledgments.}
This work was partly supported by UK Defence Science and Technology Labs under Centre for Defence Enterprise grant CDE36620. We would like to thank Dr Georgios Giasemidis for useful discussions and his help with preprocessing of the data.
\clearpage

\bibliographystyle{abbrv}
\bibliography{sigproc1}

\appendix

\section{The data we are making available}
\label{app:data_made_available}

We propose to make available the curated data sets used for the various analyses in described in this article. These will be:
\begin{itemize}
\item The seven-day evolving network used for the communicability analysis described in Section~\ref{sec:communicability_and_sent}.
\item The graph used for community detection, as described in Section~4.\ref{sub:community_detection}.
\item The tweets within each community that we collected (as described in Section~\ref{sec:data}); this data covers the analysis done in Sections~4.\ref{sub:endurance},
4.\ref{sub:dynamics_of_sentiment} and \ref{sec:abm}.
\end{itemize}
For each tweet in these data sets, we propose to include the following attributes:

\begin{enumerate}
\item an anonymised tweet ID,
\item a timestamp,
\item who was the sender (an anonymised user ID),
\item who was mentioned in the tweet (anonymised user IDs), and
\item sentiment scores for the three measures (MC), (SS) and (L).
\end{enumerate}

\section{Extracting a mentions network}
\label{app:extracting_network}

\begin{figure}[t]
\centering
\includegraphics[width=0.9\textwidth]{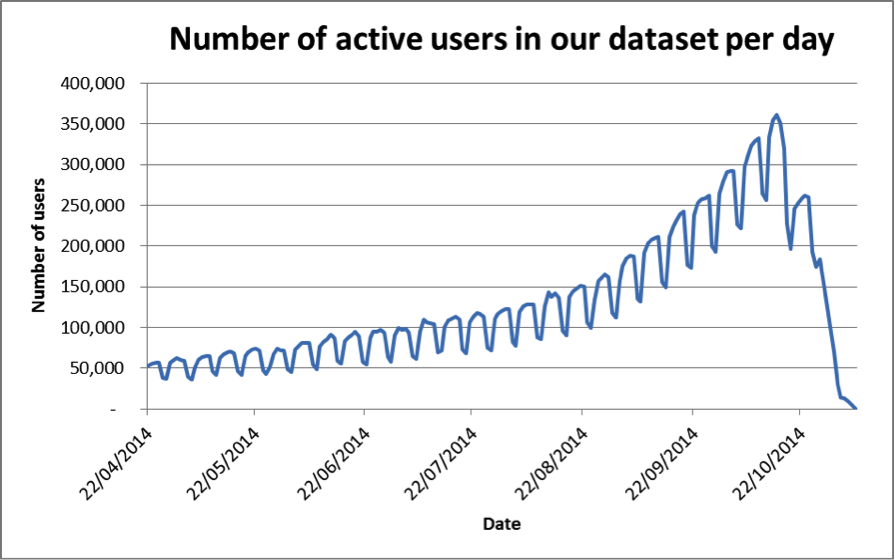}
\caption{Number of active users per day. Dates before 22nd April 2014 have fewer than 53,000 active users.}
\label{fig:no_active_users}
\end{figure}

\begin{figure}[t]
\centering
\includegraphics[width=0.9\textwidth]{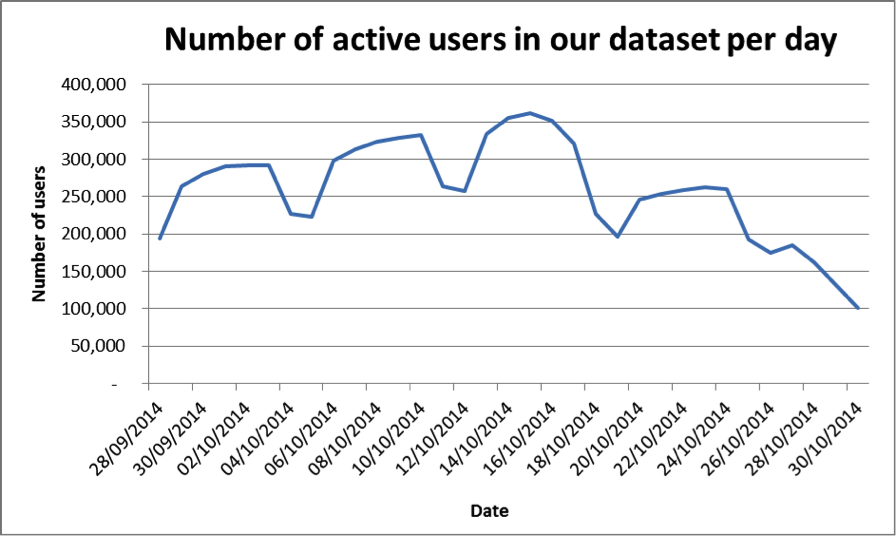}
\caption{Number of active users per day for a restricted range of dates.}
\label{fig:no_active_users1}
\end{figure}

To get the best results, we have chosen for analysis the period with the largest possible number of users active in our data. Fig.~\ref{fig:no_active_users} shows the number of users active in the data each day for the period from 22nd April 2014 until the end of the snowball-sampled data. Fig.~\ref{fig:no_active_users1}  shows the same thing but ``zoomed in'' to a restricted range of dates.
We notice that the number of users oscillates between weekdays and weekends, and the weekly total gradually increases and peaks on 15th October 2014. Then the number of users falls off quite rapidly. The shape of Fig.~\ref{fig:no_active_users} is largely due to the fact that only the last 200 tweets (from the time of the API request) per user were collected. Thus those users who tweet frequently do not show up in the first half of the chart since their earlier tweets have not been collected.

We used the Twitter dataset to create an evolving network, where the set of vertices is fixed and the edges between them can change in each time-step (a day in our case). In order to choose a fixed set of vertices, we have chosen the week from 9th October to 15th October 2014 inclusive, which is the week with the highest activity measured by the number of tweets.

\begin{figure}[t]
\centering
\includegraphics[width=\textwidth]{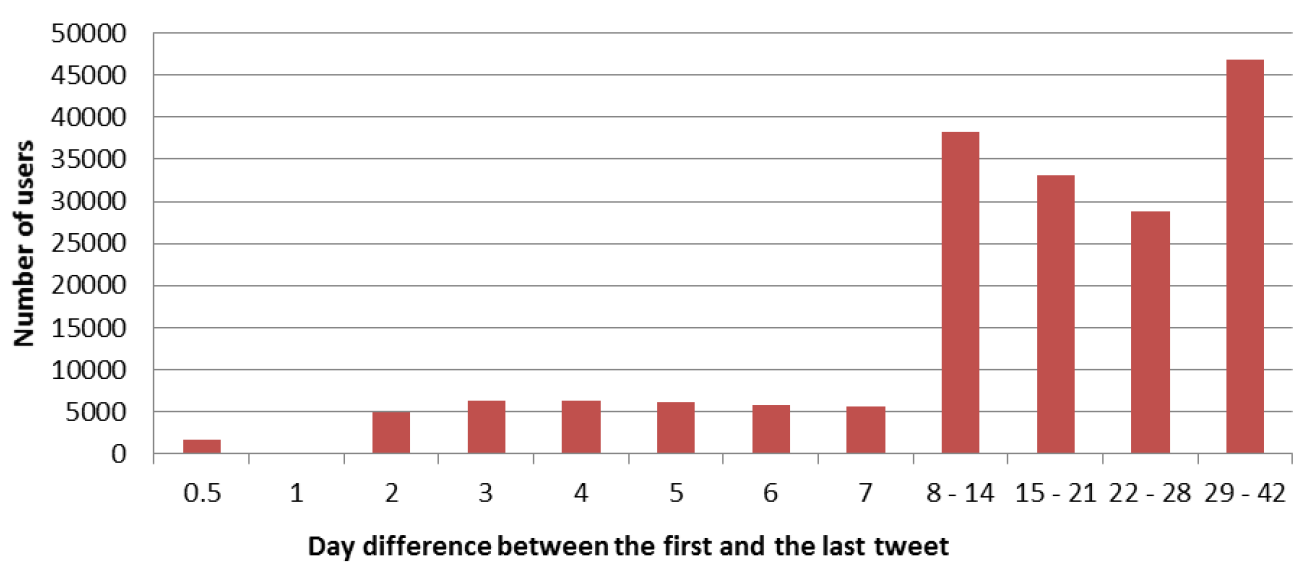}
\caption{Histogram of the number of days between the first tweet and the last tweet in our snowball-sampled data, for those users who had posted at least 200 tweets since account creation.}
\label{fig:day_differences}
\end{figure}

\begin{figure}[t]
\centering
\includegraphics[width=\textwidth]{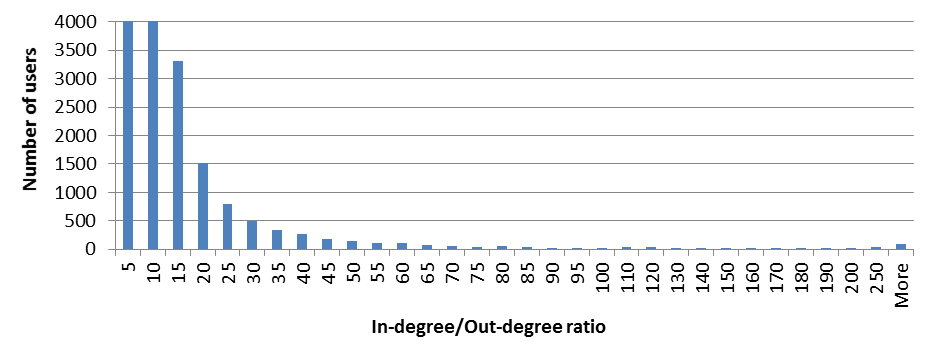}
\caption{In-degree to out-degree ratio. The first two bars have been truncated to zoom in on values greater than 10. The bin-range starts from the value of the previous bin (exclusive) up to the value under the bin (inclusive). The increase in the bin height of the last two bins is due to the increased range of the bin which includes all users with ratio from 200 to 250 and greater than 250 respectively.}
\label{fig:degree_ratios}
\end{figure}

We then filtered the data using several criteria in order to focus on ``regular'' human users. A number of classes of user with unusual behaviour were filtered out, as they would skew the results of our analyses (and threaten to make the network structure become degenerate, as we described in
Section~2.\ref{sub:extracting_mentions_network}):

\begin{itemize}

\item	Users with a very high tweeting frequency. If a user tweets hundreds of tweets in a few hours then these messages might have been automatically generated. This practice is followed by many companies and organisations for advertising purposes, but the messages are not genuinely representative of human behaviour.
Fig.~\ref{fig:day_differences} shows the number of days (rounded up to the next value) between the first tweet and the last tweet in our snowball-sampled data, just for those users who had posted at least 200 tweets since account creation.
When setting a threshold on tweeting frequency to exclude users, we should of course filter out only a small minority of users. We observe in Fig.~\ref{fig:day_differences} a natural ``gap'' in the data at the value 1, with only 48 users appearing in this bin. We select a day difference of 1 --- equivalently a tweeting frequency of 200 tweets per day --- as our threshold, excluding 1,153 users with a higher tweeting frequency than this.

\item	Users who mention themselves very frequently, who may also be bots. We used a threshold of 0.5 for the ratio of the number of self-mentions to the number of all mentions made by a user, chosen because the number of users with a self-mention ratio larger than $R$ drops off rapidly as $R$ increases above 0.5. The 0.5 threshold also seems to be a reasonable choice because it indicates that outliers mention themselves more often than they mention all other users.

\item	Users with a high ratio of in-degree to out-degree. Examples of these users are celebrities or well-known services which attract a high number of mentions relative to their activity. Looking at Fig.~\ref{fig:degree_ratios}, we observe that the number of users smoothly decreases as the in-degree to out-degree ratio increases. Since there is no value beyond which the number of users drastically decreases, there is no clear choice of threshold. We set the threshold at 50, meaning we treat as outliers, and exclude, users with in-out ratio greater than 50. In other words, we assume that users that receive mentions 50 times more than they mention others are celebrities/politicians or big organisations that skew the network and should be excluded. Indeed among the users with exceptionally high ratio one can find
TheEconomist, UberFacts, MayorofLondon, amandabynes, NatGeo, HillaryClinton, Ed\_Miliband, BBCPanorama, David\_Cameron, JunckerEU, BillGates  and YouTube.
\end{itemize}
After these filtering steps 304,349 users remained.

We wanted our evolving network to reflect users' \emph{conversations}, rather than one-way messaging, so we performed one more filtering step. We formed an undirected network on the remaining users by using only \emph{reciprocated} mentions; this means that we put an edge between users $A$ and $B$ just when $A$ had mentioned $B$ sometime during the chosen week and also $B$ had mentioned $A$ during the chosen week. Then we found the largest connected components of this graph, which contained 285,168 users (i.e.\ 94\% of the 304,349 users). We took these 285,168 users as our final set of nodes; they form a ``proper'' social network in the sense that there is a path of reciprocal mentions connecting any pair of users.

We emphasise that the reciprocal mentions as undirected edges were only used for choosing the final node set; the seven one-day snapshots that formed the evolving network we studied did include all the mentions between the chosen users, even unrecriprocated ones.

\section{ABM global parameters and rules}
\label{app:abm_params_and_rules}

Here we describe in more detail the global parameters of our agent-based model, and the rules governing the agents' behaviour.
The six global parameters are as follows:
\begin{itemize}

\item Number of iterations (discrete time steps) per day.

\item Mean number of messages per burst ($\mathit{MeanBurstSize}$). When an agent in the model decides to send something to another agent, it will issue a burst of one or more messages together. This reflects the fact that tweets are limited in length, so sometimes a quick succession of tweets is needed to convey a thought. This parameter sets the mean number of messages in these bursts.

\item Contagion of sentiment factor ($\mathit{ContagionFactor}$). There is evidence of contagion of emotion through social networks (see e.g.\ \cite{Kramer:2014}). This means that users' moods are raised when they receive positive messages and lowered when they receive negative messages. This parameter controls the extent to which an agent's sentiment is affected by the sentiment of the messages it receives.

\item Sentiment reset probability ( $P(reset)$ ). We observed that users' sentiments tend to fluctuate around a (user-specific) baseline level, and that showing sentiment higher or lower than this baseline level does not ``carry over'' to the next day. To make sure that the sentiments of our agents do not carry over either, there is a chance in each iteration that the agent's sentiment will randomly reset to that agent's baseline sentiment level. This parameter controls the probability of such a reset.

\item Sentiment noise level ($\mathit{SentimentNoiseLevel}$). Although they have a baseline sentiment, users do not post every tweet with exactly the same sentiment; there is some variation or noise around the baseline. This parameter controls the amount of that noise or variation.

\item Neighbour frequency threshold. This controls which agents will be set up as neighbours in the model. If the neighbour threshold is 10, for example, then only users who have exchanged at least 10 messages (in either direction) will be connected in the model's graph. The threshold is included so that, if desired, we can make sure that only regular correspondents will be made neighbours in the model. Note that this parameter does not affect the operation of the model, but only the creation of the model from the historical data.
\end{itemize}
Let us now explain in detail how agents decide when to send messages, and how the sentiment of each agent evolves over time. The rules governing the sending of messages are as follows.
\begin{itemize}
\item	If the agent $A$ has received messages from any other agents in the last time step, $A$ will decide whether or not to reply to these agents, and whether or not to propagate to its other neighbours. For each agent $B$ who sent $A$ a message, $A$ will reply with probability $P(reply, A)$. For each neighbouring agent $C$ who did not sent $A$ a message, $A$ will propagate a message with probability $P(prop, A)$.

\item	If agent $A$ received no messages in the previous time step, $A$ will decide whether or not to initiate a conversation with its neighbours. For each neighbouring agent $B$, $A$ will initiate a conversation with $B$ with probability $P(init, A)$.

\item	When an agent $A$ chooses to initiate, reply or propagate to another agent, it sends a burst of $n+1$ messages with $n$ drawn from a Poisson distribution with mean $\mathit{MeanBurstSize} - 1$ (this ensures a minimum burst size of 1).

\item	When an agent $A$ chooses to initiate, reply or propagate to another agent, the sentiment of the messages is generated by taking the agent's current sentiment level and adding Gaussian noise with standard deviation $\mathit{SentimentNoiseLevel}$. The resulting values are capped to the appropriate range: $-25$ to +25 for (MC), $-4$ to $4$ for (SS) and
$-100$ to 100 for (L). When working with the (MC) and (SS) sentiment measures, which are integer-valued, the values are also rounded to the nearest integer.

\end{itemize}
The rules that govern how each agent's sentiment evolves from one time step to the next are as follows.
\begin{itemize}
\item	With probability $P(reset)$  the agent's sentiment level is reset to the agent's baseline level $S(baseline, A)$.

\item	Otherwise, the agent's current sentiment level has a component added to it representing the influence of each message received in the last time step. The component for a message received with sentiment $S$  is
$(S - S(neutral, A)) \times \mathit{ContagionFactor}$.
\end{itemize}

\section{Parameter space for simulation runs}
\label{app:param_space}

\begin{description}
\item {\bf Number of iterations per day}. In theory this parameter could be set to any integer value. However, increasing this parameter also increases the execution time of simulations, which enforces a limit in practice. By starting from 24 iterations per day and successively doubling, we arrived at a limit of 1,536 iterations per day (that is, $24 \times 2^6$). Values above this required too much processing time to be practical.

\item	{\bf Mean number of messages per burst}. This parameter has a lower bound of 1. We examined in the real data the numbers of times each user $A$ sent multiple messages to another user $B$ within a period of $t$ seconds, for various values of $t$. We decided to test values in the range from $1.1$ to $2.8$.

\item	{\bf Neighbour threshold}. We tested values from 1 to 60. Setting the threshold to 60 makes the graph very sparse, yielding only 77 edges among the 28 users.

\item	 {\bf Contagion of sentiment factor}. A value of 1 for this parameter would mean that when a user receives a message, the sentiment of that one message is approximately just as important to the user's future sentiment as the user's entire history to date. Thus the value 1 seems implausibly high. We tested values from the lower bound of 0 up to 0.5.

\item	{\bf Sentiment reset probability}. This parameter naturally ranges from 0 to 1, but we tested only values from 0 to 0.5 for the following reason. Values greater than 0.5 cause the sentiment to be reset on the majority of iterations, which means that users' sentiment levels never move far from their baseline levels. But this situation is already covered by the case when the contagion of sentiment factor is zero.

\item	{\bf Sentiment noise level}. The standard deviation of the (MC) sentiment scores of messages sent within the studied community 17 was 1.57, so we knew that values of the sentiment noise level parameter much larger than this would not perform well. Thus for (MC) we tested values from the lower bound of 0 up to 2.5. By similar reasoning, we selected the range from 0 to 1.8 for (SS) and from 0 to 13 for (L).

\end{description}

\end{document}